\documentclass[twocolumn]{aastex7}
\shorttitle{51~Eri~b Interferometry}
\shortauthors{Balmer et al.}
\definecolor{mylinkcolor}{RGB}{168, 63, 87}

\hypersetup{linkcolor=mylinkcolor,
citecolor=mylinkcolor,
urlcolor=mylinkcolor}

\usepackage{xspace}
\usepackage{amsmath}
\usepackage{subcaption}

\newcommand{\um}{\text{\,\textmu m}\xspace}
\newcommand{\Mjup}{M_{\rm{Jup}}\xspace}

\begin{document}

\title{On the High Eccentricity Orbit of 51~Eri~b and Constraints on Interactions with a Second Companion}

\suppressAffiliations

\author[0000-0001-6396-8439]{William O. Balmer}
\altaffiliation{51 Pegasi b Fellow}
\affiliation{Center for Interdisciplinary Exploration and Research in Astrophysics (CIERA), Northwestern University, 1800 Sherman Avenue, Evanston, IL 60201, USA}
\email{balmer@northwestern.edu}
\correspondingauthor{Will Balmer, \href{mailto:balmer@northwestern.edu}{balmer@northwestern.edu}}

\author[0000-0001-5684-4593]{William Thompson}
\email{william.thompson@nrc-cnrc.gc.ca}
\affiliation{NRC Herzberg Astronomy and Astrophysics,
5071 West Saanich Road,
Victoria, BC, V9E 2E7, Canada}

\author[0009-0008-5322-9219]{Rodrigo P. Silva}
\email{rpdasilva17@gmail.com}
\affiliation{CFisUC, Departamento de Física, Universidade de Coimbra, 3004-516 Coimbra, Portugal}
\affiliation{Faculdade de Engenharia, Universidade do Porto, Rua Dr. Roberto Frias, 4200-465 Porto, Portugal}
\affiliation{CENTRA – Centro de Astrofísica e Gravitação, IST, Universidade de Lisboa, 1049-001 Lisboa, Portugal}

\author[0000-0003-3818-408X]{Laurent Pueyo}
\email{pueyo@stsci.edu}
\affiliation{Space Telescope Science Institute, 3700 San Martin Drive, Baltimore, MD 21218, USA}

\author[0000-0002-6948-0263]{Sylvestre Lacour}
\email{sylvestre.lacour@obspm.fr}
\affiliation{LESIA, Observatoire de Paris, PSL, CNRS, Sorbonne Universit\'e, Universit\'e de Paris, 5 place Janssen, 92195 Meudon, France}
\affiliation{European Southern Observatory, Karl-Schwarzschild-Str. 2, 85748 Garching, Germany}


\author[0000-0002-8946-8579]{Alexandre C. M. Correia}
\email{acm.correia@gmail.com}
\affiliation{CFisUC, Departamento de Física, Universidade de Coimbra, 3004-516 Coimbra, Portugal}
\affiliation{LTE, Observatoire de Paris, Université PSL, Sorbonne Université, CNRS, 75014 Paris, France}

\author{Mathias Nowak}
\email{mathias.nowak@obspm.fr}
\affiliation{LIRA, Observatoire de Paris, PSL, CNRS, Sorbonne Universit\'e, Universit\'e de Paris, 5 place Janssen, 92195 Meudon, France}

\author[0000-0001-9582-4261]{Dori Blakely}
\email{blakelyd@uvic.ca}
\affiliation{Department of Physics and Astronomy, University of Victoria,
3800 Finnerty Road, Elliott Building,
Victoria, BC, V8P 5C2, Canada}
\affiliation{NRC Herzberg Astronomy and Astrophysics,
5071 West Saanich Road,
Victoria, BC, V9E 2E7, Canada}

\author[0000-0002-3191-8151]{Marshall D. Perrin}
\email{mperrin@stsci.edu}
\affiliation{Space Telescope Science Institute, 3700 San Martin Drive, Baltimore, MD 21218, USA}

\author{Allan Denis}
\email{allan.denis@lam.fr}
\affiliation{Aix Marseille Univ, CNRS, CNES, LAM, Marseille, France}

\author[0000-0003-0774-6502]{Jason J. Wang}
\email{jason.wang@northwestern.edu}
\affiliation{Center for Interdisciplinary Exploration and Research in Astrophysics (CIERA), Northwestern University, 1800 Sherman Avenue, Evanston, IL 60201, USA}
\affiliation{Department of Physics and Astronomy, Northwestern University, 2145 Sheridan Road, Evanston, IL 60208, USA}

\author[0000-0003-2769-0438]{Jens Kammerer}
\email{jens.kammerer@eso.org}
\affiliation{European Southern Observatory, Karl-Schwarzschild-Str. 2, 85748 Garching, Germany}

\author[0000-0002-1678-3535]{Paulo J. V. Garcia}
\email{pgarcia@fe.up.pt}
\affiliation{Faculdade de Engenharia, Universidade do Porto, Rua Dr. Roberto Frias, 4200-465 Porto, Portugal}
\affiliation{CENTRA – Centro de Astrofísica e Gravitação, IST, Universidade de Lisboa, 1049-001 Lisboa, Portugal}

\author[0000-0003-1278-4722]{Jonathan Roberts}
\email{jonathanroberts2027@u.northwestern.edu}
\affiliation{Center for Interdisciplinary Exploration and Research in Astrophysics (CIERA), Northwestern University, 1800 Sherman Avenue, Evanston, IL 60201, USA}
\affiliation{Department of Physics and Astronomy, Northwestern University, 2145 Sheridan Road, Evanston, IL 60208, USA}

\author{Guillaume Bourdarot}
\email{bourdarot@mpe.mpg.de7}
\affiliation{Max-Planck-Institut f\"{u}r Astronomie, K\"onigstuhl 17, 69117 Heidelberg, Germany} 

\author{Ga\"el Chauvin}
\email{chauvin@mpia.de}
\affiliation{Max-Planck-Institut f\"{u}r Astronomie, K\"onigstuhl 17, 69117 Heidelberg, Germany} 
\affiliation{Laboratoire Lagrange, Universite\' Cote d'Azur, CNRS, Observatoire de la Cote d'Azur, 06304 Nice, France}

\author{Thomas Henning}
\email{henning@mpia.de}
\affiliation{Max-Planck-Institut f\"{u}r Astronomie, K\"onigstuhl 17, 69117 Heidelberg, Germany}

\author[0000-0001-8074-2562]{Sasha Hinkley}
\email{S.Hinkley@exeter.ac.uk}
\affiliation{Department of Physics \& Astronomy, University of Exeter, Physics Building, Stocker Road, Exeter EX4 4QL, UK}

\author[0000-0002-9803-8255]{Kielan K. W. Hoch}
\email{khoch@stsci.edu}
\affiliation{Space Telescope Science Institute, 3700 San Martin Drive, Baltimore, MD 21218, USA}

\author[0000-0003-0626-1749]{Pierre Kervella}
\email{pierre.kervella@obspm.fr}
\affiliation{LIRA, Observatoire de Paris, Universit\\'e PSL, Sorbonne Universit\'e, Universit\'e Paris Cit\'e, CY Cergy Paris Universit\'e, CNRS, 92190
Meudon, France}
\affiliation{French-Chilean Laboratory for Astronomy, IRL 3386, CNRS and U. de Chile, Casilla 36-D, Santiago, Chile}

\author[0000-0003-0514-1147]{Laura Kreidberg}
\email{kreidberg@mpia.de}
\affiliation{Max-Planck-Institut f\"ur Astronomie, K\"onigstuhl 17, D-69117 Heidelberg, Germany}

\author[0000-0001-7827-7825]{Roeland P. van der Marel}
\email{marel@stsci.edu}
\affiliation{Space Telescope Science Institute, 3700 San Martin Drive, Baltimore, MD 21218, USA}
\affiliation{Department of Physics \& Astronomy, Johns Hopkins University, 3400 N. Charles Street, Baltimore, MD 21218, USA}

\author[orcid=0000-0002-2918-8479,sname='Mathilde Mâlin']{Mathilde Mâlin} 
\email{mmalin@stsci.edu}
\affiliation{Department of Physics \& Astronomy, Johns Hopkins University, 3400 N. Charles Street, Baltimore, MD 21218, USA} 
\affiliation{Space Telescope Science Institute, 3700 San Martin Drive, Baltimore, MD 21218, USA}

\author[0000-0002-9765-7783]{Florentin Millour}
\email{florentin.millour@oca.eu}
\affiliation{Universit\'e C\^ote d'Azur, Observatoire de la C\^ote d'Azur, CNRS, Laboratoire Lagrange, France}

\author[0009-0003-4601-8556]{Nitya Pandey}
\email{npandey@das.uchile.cl}
\affiliation{Departamento de Astronom\'ia, Universidad de Chile, Camino del Observatorio 1515, Las Condes, Santiago, Chile}
\affiliation{European Southern Observatory, Alonso de Córdova 3107, Vitacura, Santiago, Chile}

\author[0000-0003-4203-9715]{Emily L. Rickman}
\email{erickman@stsci.edu}
\affiliation{European Space Agency (ESA), ESA Office, Space Telescope Science Institute, 3700 San Martin Drive, Baltimore, MD 21218, USA}

\author[0000-0001-9992-4067]{Matthias Samland}
\email{samland@mpia.de}
\affiliation{Max-Planck-Institut f\"{u}r Astronomie, K\"{o}nigstuhl 17, 69117 Heidelberg, Germany}

\author[0000-0002-5823-3072]{Tomas Stolker}
\email{stolker@strw.leidenuniv.nl}
\affiliation{Leiden Observatory, Leiden University, Einsteinweg 55, 2333 CC Leiden, The Netherlands}

\author[0000-0002-5902-7828]{Arthur Vigan}
\email{arthur.vigan@lam.fr}
\affiliation{Aix Marseille Univ, CNRS, CNES, LAM, Marseille, France}

\author[0000-0002-6618-1137]{Jerry W. Xuan}
\altaffiliation{51 Pegasi b Fellow}
\email{jerryxuan@g.ucla.edu}
\affiliation{Department of Earth, Planetary, and Space Sciences, University of California, Los Angeles, CA 90095, USA}




\begin{abstract}

The eccentricity of an exoplanet's orbit is a key observational quantity that can be used to infer its dynamical past and predict its future orbital evolution. Planets are expected to emerge from their protoplanetary cradles with roughly circular eccentricities, and non-circular eccentricites to arise from subsequent dynamical processing in the absence of the gas disk. We present an updated orbit for 51~Eridani~b based on 5 years of 0.2 mas precision optical interferometry from VLTI/GRAVITY(+). This planet is young (${\sim}20\,\mathrm{Myr}$ old), temperate ($T_{\rm eff}\sim650\,\mathrm{K}$), and on a wide orbit ($a{=}9.36\,\mathrm{au}$) with an uncertain mass ($M_{\rm b}\leq12\,M_{\rm J}$).
We measured the planet's motion to an exceptionally high precision, confirming that its orbit is very eccentric ($e_b=0.574\pm0.003$). We observe residuals to the absolute astrometry of the host star based on a single planet orbit that could be attributable to reflex motion from a third body in the system. We construct a two companion orbit model that could explain the apparent reflex motion of 51~Eri~A: this model predicts that the putative second companion has $M_c\in[1.0, 2900]_{95\%}\,M_{\rm Jup}$ and $a_c\in[0.3, 3.1]_{95\%}\,\mathrm{au}$. In this framework, a dynamical interaction between b and ``c" could have kicked the exterior, lower mass b out onto its present, eccentric orbit. If this signal is astrophysical, and not due to a systematic error, this candidate should be confirmable with Gaia DR4 epoch astrometry in the near future.
\end{abstract}

\section{Introduction} \label{sec:intro}
\par Gas giants are the dominant dynamical forces in their planetary systems, and are therefore crucial to understanding our own Solar System in a broad context. The distribution of giant planet eccentricity contains clues regarding giant planet formation and dynamical evolution \citep[e.g.][]{Zakamska2004, Ford2008, Bowler2020, Blunt2026}. The eccentricities of giant planets detected via radial velocities exhibit a large population of low eccentricity orbits, with a tail towards higher eccentricities \citep{Kipping2013, Blunt2026, Morgan2026}. Near-circular orbits are the expected result of formation within a protoplanetary disk, where gas drag damps the initial eccentricity. Subsequent processing, by scattering with other giant planets, interactions with stellar multiples, or stellar close encounters is then expected to increase the eccentricities for a fraction of these systems. Wider separation, directly imaged companions, when sorted by mass, appear to exhibit a preference for low eccentricity orbits \citep{Bowler2020}, although this apparent trend is complicated by the low sample size, choice of priors (and hyper-priors), and limited coverage of orbital arcs \citep{DoO2023, Nagpal2023}.

\par 51~Eridani~b is one of the lowest mass directly imaged planets known. It was discovered during the Gemini Planet Imager (GPI) Exoplanet Survey \citep{Macintosh2015} and has been subsequently characterized by Gemini/GPI \citep{DeRosa2015, Rajan2017, DeRosa2020}, VLT/SPHERE \citep{Samland2017, Maire2019, Brown-Sevilla2023}, Keck/NIRC2 \citep{Rajan2017, Madurowicz2023}, JWST/NIRCam \citep{Balmer2025b}, VLT/HiRISE \citep{Denis2026}, and JWST/NIRSpec \citep{Madurowicz2025}. Atmospheric modeling has indicated that the planet has a cloudy atmosphere in chemical disequilibrium due to vertical mixing \citep{Rajan2017, Brown-Sevilla2023, Madurowicz2023, Whiteford2023, Balmer2025b, Madurowicz2025}. Given the planet's low luminosity and the system's apparent age \citep{Elliott2024} based on its membership in the $\beta$-Pic Moving Group (BPMG) \citep[e.g.][]{Zuckerman2001, Bell2015}, the planet's mass has been projected to lie below the deuterium burning limit \citep{Macintosh2015}. \added{The proper motion anomaly between \textit{Hipparcos} and \textit{Gaia} for the host star is incongruent with the reflex motion that the planet's orbit should induce at $1{-}2\,\sigma$ \citep[see Figure 5-9 in][]{DeRosa2020}. When this proper motion anomaly is included in the likelihood function of a single planet orbit fit, it implies an upper limit on the mass of the planet of ${\leq}12\,\Mjup$ at $2\,\sigma$ \citep{Dupuy2022}. }

\par By tracing the visual motion of 51~Eri~b over 10 years, direct imaging observations have indicated preference for a high eccentricity orbit ($e\simeq0.6$), albeit with skewed distributions of $e$ that still encompass circular orbits at lower probability \citep[][]{DeRosa2015, Maire2019, DeRosa2020, Dupuy2022, Balmer2025b, Denis2026}. The planet appears to be a notable exception to the apparent clustering of low eccentricities for planetary mass, directly imaged companions. In this paper we seek to confirm the high eccentricity of 51~Eri~b, and to place constraints on the processes that could have excited the planet on to such an orbit. 

\section{Observations} \label{sec:obs}

\par We observed 51~Eri~b 14 times between 2019 and 2025, as part of the ExoGRAVITY Large Program (1104.C-0651, PI: S.~Lacour), the JWST Telescope Scientist Team (JWST-TST)\footnote{\url{https://www.stsci.edu/~marel/jwsttelsciteam.html}} Guaranteed Time Observations (PI: M.~Mountain, subject lead: M.~Perrin), a dedicated VLTI Monitoring Program (111.24MN, PI: W.~Balmer), and the commissioning of GRAVITY+ \citep{gravity_plus_firstlight}. Table \ref{tab:obslog} records the observing log of successful detections of the planet.

\begin{deluxetable*}{lccclllc}
\tablewidth{\textwidth}
\tablecaption{New observations of 51~Eri described in this work.\label{tab:obslog}}
\tablehead{Date & \multicolumn{2}{c}{Target} & NEXP/NDIT/DIT & Airmass & $\tau_0$ & Seeing & Fiber pointing \\
UTC & SC/MASK & FT/NGS & & & ms & $''$ & mas}
\startdata
\hline\hline
SPHERE & & & & & & & \\
2021-10-28$^*$     & CLEAR  & 51~Eri & 9/25/24~s    & 1.08-1.18& 5 & 0\farcs6-1\farcs2 & \nodata \\
2021-10-28$^*$     & N\_ALC\_Ks & 51~Eri,  & 2/100/0.837~s& 1.08-1.18& 5 & 0\farcs6-1\farcs2 & \nodata \\
\hline
GRAVITY & & & & & & & \\
2022-01-23$^*$$^*$ & 51~Eri~A     & 51~Eri~A     & 6/64/1~s    & 1.14-1.15 & 6.3-7.5  & 0\farcs-1\farcs05   & (0,0) \\
2022-01-23$^*$$^*$ & 51~Eri~b     & 51~Eri~A     & 14/4/100~s  & 1.14-1.73 & 5.9-20.4 & 0\farcs27-1\farcs31 & (291.9, -218.3) \\
2022-01-23$^*$$^*$ & HD~73900~A/B & HD~73900~B/A & 4/64/1~s    & 1.04-1.05 & 6.4-19.1 & 0\farcs27-0\farcs67 & (-1108.2, 710.2) \\
2022-09-15$^\dag$  & HD~25535~A/B & HD~25535~B/A & 4/64/1~s    & 1.11-1.13 & 3.5-3.7  & 0\farcs57-0\farcs60 & (752.5, 787.1) \\
2022-09-15$^\dag$  & 51~Eri~b     & 51~Eri~A     & 14/4/100~s  & 1.08-1.29 & 1.7-2.8  & 0\farcs76-1\farcs10 & (294.2, -177.9) \\
2022-09-15$^\dag$  & 51~Eri~A     & 51~Eri~A     & 6/64/1~s    & 1.08      & 2.4-3.2  & 0\farcs69-0\farcs86 & (0,0) \\
2024-11-15$^\ddag$ & 51~Eri~A     & 51~Eri~A     & 6/64/1~s    & 1.08-1.56 & 6.7-10.7 & 0\farcs40-0\farcs85 & (0,0) \\
2024-11-15$^\ddag$ & 51~Eri~b     & 51~Eri~A     & 14/4/100~s  & 1.08-1.71 & 6.7-10.7 & 0\farcs40-0\farcs85 & (250.8, -14.2) \\
2024-12-15$^\S$    & 51~Eri~A     & 51~Eri~A     & 6/64/1~s    & 1.08-1.21 & 4.0-12.4 & 0\farcs49-0\farcs77 & (0,0) \\
2024-12-15$^\S$    & 51~Eri~b     & 51~Eri~A     & 14/4/100~s  & 1.08-1.21 & 4.0-12.4 & 0\farcs49-0\farcs77 & (263.5, -17.8) \\
2025-10-03$^\S$    & 51~Eri~A     & 51~Eri~A     & 6/64/1~s    & 1.08-1.27 & 2.1-5.5  & 0\farcs49-0\farcs81 & (0,0) \\
2025-10-03$^\S$    & 51~Eri~b     & 51~Eri~A     & 14/4/100~s  & 1.08-1.27 & 2.1-5.5  & 0\farcs49-0\farcs81 & (230.3, 37.5) \\
\hline
JWST & & & & & & & \\
2022-10-19 & 430ROUND & 51~Eri & 2/41/DEEP8 & 0.0 & \nodata & \nodata & \nodata
\enddata
\tablecomments{$*$ 107.22RM.001, $**$ 1104.C-0651(B),  $\dag$ 1104.C-0651(F), $\ddag$ G+ technical time, $\S$ 111.24MN. For GRAVITY dual-field observations, Science Camera (SC) and Fringe Tracker (FT) targets are recorded separately. For the SPHERE observation the coronagraphic mask and the natural guide star (NGS) target are recorded instead. NEXP, NDIT, and DIT denote the number of exposures, the number of detector integrations per exposure, and the detector integration time, respectively. $\tau_0$ denotes the atmospheric coherence time. The fiber pointing is the placement of the science fiber relative to the fringe tracking fiber (which is placed on the central star).
}
\end{deluxetable*}

\subsection{VLT/SPHERE Observations}

51~Eri was observed with the Spectro-Polarimetric High-contrast Exoplanet REsearch instrument \citep[SPHERE,][]{Beuzit2019} on 2021-10-28, simultaneously using the Infra-Red Dual-bean Imaging and Spectroscopy camera \citep[IRDIS,][]{Dohlen2008} and the Integral Field Spectrograph \citep[IFS,][]{Claudi2008} at the European Southern Observatory's Very Large Telescope at the Paranal Observatory in Chile. Observations consisted of a flux and Point Spread Function (PSF) calibration series taken of the star through a neutral density filter, and then of coronagraphic observations taken in pupil-tracking mode to facilitate angular differential imaging (ADI). Each coronagraphic observation had satellite spots, \added{copies of the stellar point spread function re-imaged outside of the coronagraphic mask}, applied using the deformable mirror, enabling a constant monitoring of the star's position and flux despite the presence of the coronagraph \citep{Marois2006, Sivaramakrishnan2006}. Observing conditions were acceptable, with the atmospheric seeing varying between 0\farcs6 and 1\farcs2, and on average a coherence time of $\sim5\,\mathrm{ms}$. The observing sequence had an effective field rotation of $46.7^\circ$. As we are interested in the planet's astrometry, we only consider the IRDIS dataset in this work, the astrometric solution for which was last calculated in \citet{Maire2021}.

\par The data were reduced using \texttt{vlt-sphere}\footnote{\url{https://ascl.net/2009.002}} \citep{Vigan2020} a \texttt{python} language package wrapping the ESO SPHERE pipeline v0.58.1\footnote{\url{https://www.eso.org/sci/software/pipelines/sphere/}}. Bias, dark, background, and flat field corrections were applied using calibrations taken during the same night, and individual integrations were co-added by exposure, resulting in 9 images taken in two filters, K1 (2110 nm) and K2 (2251 nm). 51~Eri~b's spectral type exhibits strong methane absorption between these two filters, so we only considered measuring astrometry from the K1 filter, where the planet can be reliably detected. The images were centered using the satellite spots, which were registered using a Gaussian fit within a 32x32 pixel box after high pass filtering. Images were shifted to the centering using spline interpolation.

\par
\begin{figure*}
\begin{subfigure}[c]{0.48\textwidth}
\includegraphics[width=\linewidth]{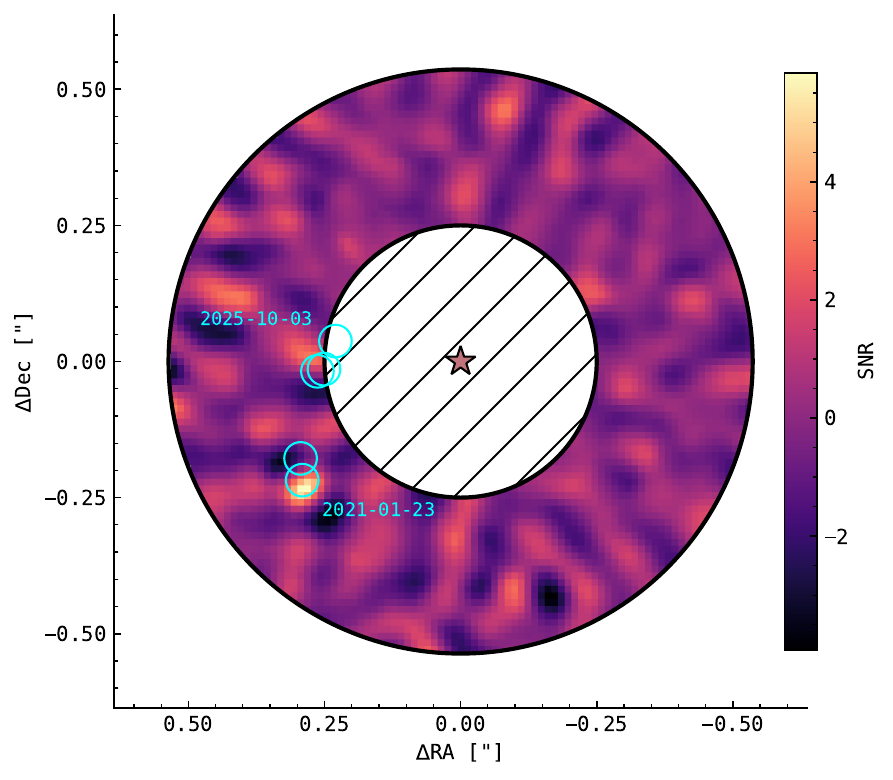} 
\caption{SPHERE, 2021-10-28}
\label{fig:subim1}
\end{subfigure}
\hfill
\begin{subfigure}[c]{0.4\textwidth}
\includegraphics[width=\linewidth]{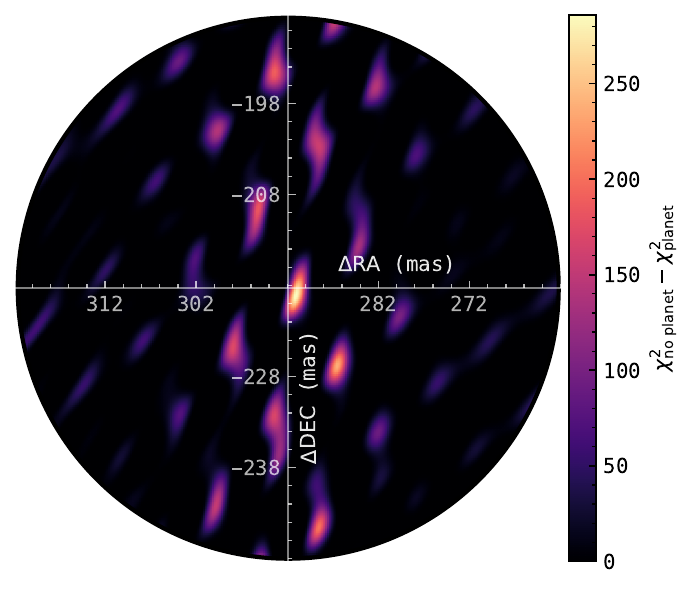}
\caption{GRAVITY, 2022-01-23}
\label{fig:subim2}
\end{subfigure}
\hfill
\begin{subfigure}[c]{0.4\textwidth}
    \hspace*{1.4cm}
    \includegraphics[width=\linewidth]{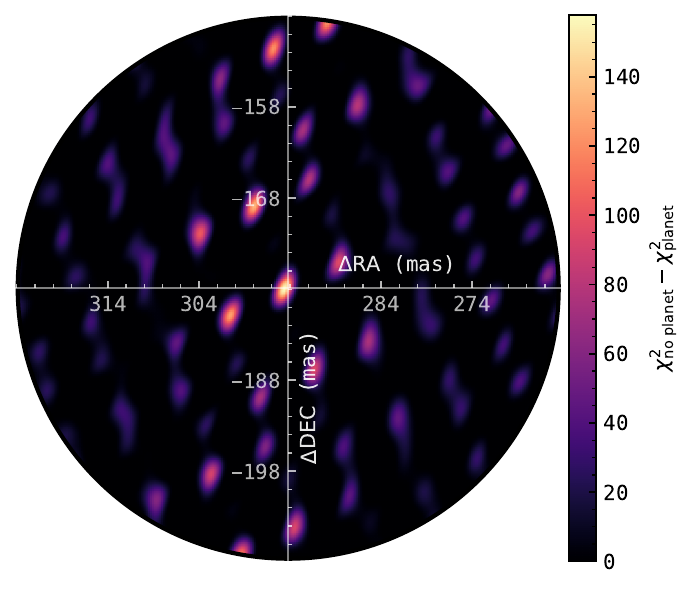}
    \caption{GRAVITY, 2022-09-15}
\label{fig:subim3}
\end{subfigure}
\hfill
\begin{subfigure}[c]{0.4\textwidth}
    \includegraphics[width=\linewidth]{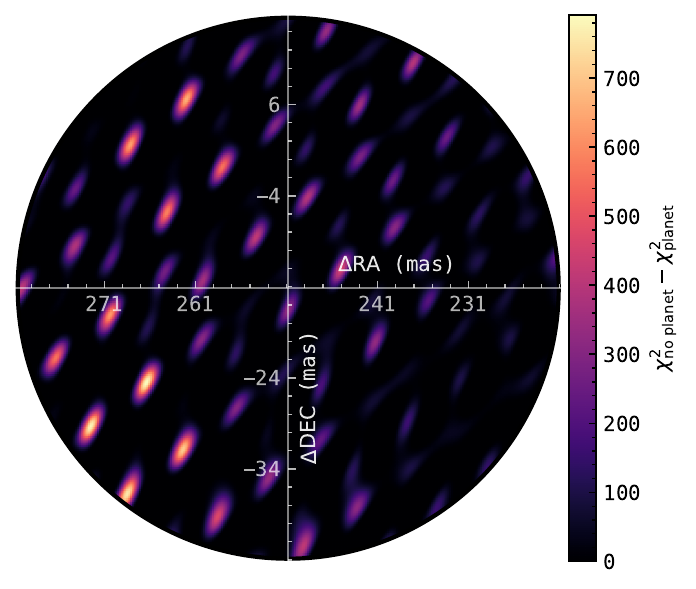}
    \caption{GRAVITY+, 2024-11-15}
\label{fig:subim4}
\end{subfigure}
\begin{subfigure}[c]{0.4\textwidth}
    \hspace*{1.4cm}
    \includegraphics[width=\linewidth]{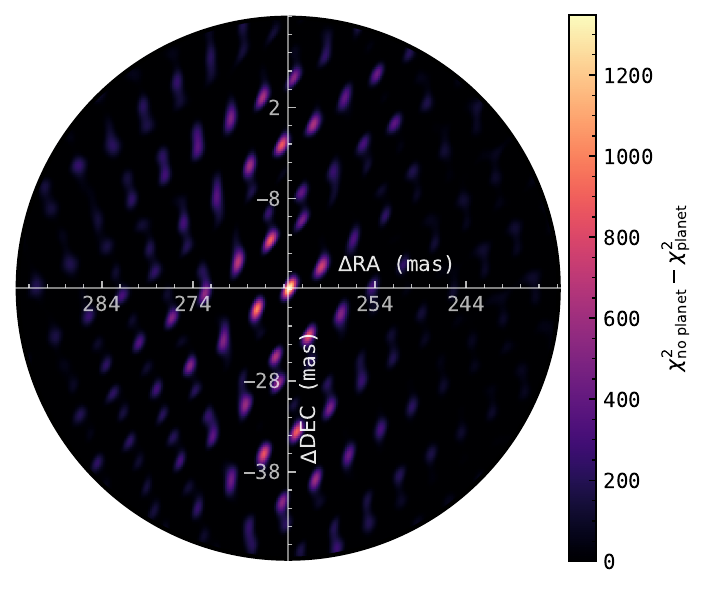}
    \caption{GRAVITY+, 2024-12-18}
\label{fig:subim5}
\end{subfigure}
\hfill
\begin{subfigure}[c]{0.4\textwidth}
    \includegraphics[width=\linewidth]{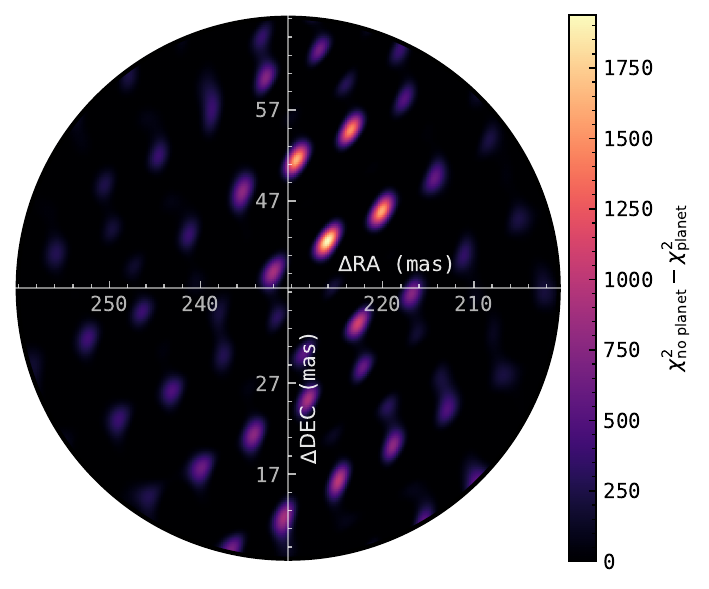}
    \caption{GRAVITY+, 2025-10-03}
\label{fig:subim6}
\end{subfigure}

\caption{Detections of 51~Eri~b. (a) SPHERE coronagraphic imaging. The planet is detected at $5\,\sigma$ in the \textit{K1}-band. Subsequent GRAVITY pointing are overlaid in light blue, illustrating the instrument field of view relative to the planetary system scene. (b-f) VLTI/GRAVITY(+) interferometry. $\chi^2_{\rm no\,\,planet}-\chi^2_{\rm planet}$ represents the likelihood the data can be described by a combination of an incoherent polynomial term capturing the stellar light leakage and instrumental/telluric effects and a coherent point source with a given on-sky position and the spectrum of a T-dwarf. This model was computed across the science fiber field of view (${\sim}60\,\mathrm{mas}$ in $\Delta$RA and $\Delta$Dec in milliarcseconds).}
\label{fig:detections}
\end{figure*}

 Residual starlight was subtracted from the IRDIS images using the Karhunen--Lo\`{e}ve Image Projection (KLIP) algorithm \citep{Soummer2012}, as implemented in the python package \texttt{pyKLIP}\footnote{\url{https://pyklip.readthedocs.io/en/latest/}} \citep{Wang2015}. We followed \citet{AdamsRedai2023} in systematically exploring KLIP parameter space to optimize the starlight subtraction at the location of the planet. As the integrations were co-added, we always used 9 principle components for constructing the PSF model, and investigated the impact of the choice of the annulus size (\texttt{annuli}) and \added{the difference in field rotation required for an image to be considered in the reference set for another image} (\texttt{movement}) parameters on the significance of the detection. Our optimization metric was the calibrated Forward Model Matched Filter \citep[FMMF,][]{Ruffio2017} Signal-to-Noise (SNR) ratio. \texttt{annuli} were defined to begin near the inner working angle at 0\farcs25 and spanned varying widths between 0\farcs15 to 0\farcs45. \texttt{movement} was varied between 0 and 7 pixels. The data strongly preferred a \texttt{movement} of 2 pixels and \texttt{annuli} between 0\farcs25-0\farcs50 and 0\farcs25-0\farcs65; these parameters yielded a SNR of 5.8 on 51~Eri~b. Our final astrometry was conducted with \texttt{movement} 2, 9 principle components, and an annulus spanning 0\farcs25-0\farcs55. This image is shown in Figure \ref{fig:subim1}.

\par Astrometry was measured via the KLIP Forward Modeling \citep{Pueyo2016} technique implemented in \texttt{pyKLIP}. The unsaturated PSF observation was used as the input forward model and degraded by a projection onto the same KLIP basis as the starlight subtracted image. The location and flux of the forward model was sampled using a Markov Chain Monte Carlo (MCMC) algorithm, implemented via the \texttt{emcee}\footnote{\url{https://emcee.readthedocs.io/en/stable/}} \citep{Foreman-Mackey2013} package, with 100 walkers, 200 discarded steps, and 800 accepted steps. We adopted the platescale uncertainty and the north angle offset uncertainty for IRDIS recorded in \citep{Maire2021}, as well as an additional star centering uncertainty of 0.25 pixels, for the astroemtric error budget. These were added in quadrature with the statistical uncertainty on the position of the planet on the detector measured from the forward model MCMC fit. This astrometry is recorded in the first row of Table \ref{tab:astrometry}, and informed the pointing of the GRAVITY Science Camera fiber in the 2021-01-23 observation.

\subsection{VLTI/GRAVITY}

51~Eri was observed many times between 2019-2025 with the GRAVITY(+) \citep{GravityCollaboration2017, gravity_plus_firstlight} on the Very Large Telescope Interferometer (VLTI) at the European Southern Observatory's Paranal Observatory in Chile. For these observations, light from the four 8.2m Unit Telescopes (UTs) is fed through fibers into the Beam Combining Instrument, where coherent light constructively interferes into fringes characteristic of the six baselines between the four telescopes (this fringe pattern can be thought of as the point spread function of the virtual telescope). \added{51~Eri~b was one of the most challenging targets to detect consistently during the course of the ExoGRAVITY large program because of its high contrast, intrinsic faintness, and late spectral type. The planet's spectrum exhibits a strong methane (CH$_4$) absorption feature between 2.2\um and 2.4\um that suppresses the planet's flux completely within observational uncertainties. This in turn reduces the effective u-v coverage of the interferometric observations by half, necessitating either ideal conditions or long observations to fill the u-v plane and achieve precise astrometry. The result of half a decade of follow-up astrometric monitoring of the planet with GRAVITY is a heterogeneous collection of observations that has required careful triage. In this work we only present astrometry derived from the subset of observations that have reliable instrument metrology and therefore reliable coherent fluxes, and that yield a significant detection of the planet. \footnote{In the future, we recommend that GRAVITY observations of mid T-type and later planets take at least $\gtrsim1.5\,\mathrm{h}$ of on-target integration to avoid similar degenerate astrometric solutions and non-detections, regardless of the planet's apparent magnitude.}}

\par We refer the reader to previous work regarding the details of the high-contrast dual field observational strategy with GRAVITY \citep[for instance,][]{GRAVITYCollaboration2019, Nowak2020, Nowak2024b, Balmer2025a}. While the planet was sufficiently separated from its host star ($>$0\farcs3), the dual-field off-axis mode was used, as this mode provides the best throughput to the Science Camera. During the dual-field off-axis observations, the metrology zero point was measured using a calibrator binary in the SWAP configuration \citep{Nowak2024b}. Once the planet moved interior to the off-axis mode limits, the dual-field on-axis mode was used; during these observations, the metrology zero point was measured using intermittent observations of the star through the Science Camera (SC) fiber. The uncalibrated data were reduced using the ESO GRAVITY pipeline v1.9.6 \footnote{\url{https://www.eso.org/sci/software/pipelines/gravity}}, up to the ``astroreduced" intermediate data-products. The subsequent phase referencing and astrometric extraction were conducted using the open source \texttt{exogravity} pipeline\footnote{\url{https://gitlab.obspm.fr/mnowak/exogravity}}. 

\par The coherent fringes of the planet are distinguished from the incoherent stellar halo by means of a polynomial and point source model comparison, as the single mode fiber and the long baseline interferometer ``resolve away" the coherent starlight. \added{The algorithm for starlight subtraction and the model for the planet's coherent emission, as implemented in the \texttt{exogravity} pipeline, is described in detail in Appendix A of \citet{Nowak2020}, and more generally in \S2.1 Eq.1-3 of \citet{Balmer2025a}.} In short, the combined point source and polynomial model is computed at each position in an arbitrarily defined grid of RA and Dec across the fiber field of view, conditioned on each of the six baseline pairs simultaneously, and repeated across each set of detector integrations to yield independent estimates of the astrometric jitter of the instrument over the course of the observation. \added{Observations with greater evidence for a detection of the planet yield a higher $\chi^2_{\rm no\,\,planet}-\chi^2_{\rm planet}$ goodness of fit metric, indicating that the data prefer the model with the coherent flux of the planet versus the starlight only model.} Importantly, we used an \texttt{Exo-REM} \citep{Charnay2018} model atmosphere as a template spectrum when describing the model of the planet's coherent flux for a given position. The model had $T_{\rm eff}=700\,{\rm K}$, $\log{g}=3.5$, ${\rm [M/H]}=0.5$, and ${\rm C/O}=0.55$ (following roughly the best fitting values in \citet{Madurowicz2025} that are closest to a grid point in the Exo-REM series), and was scaled to match the flux of the GPI \textit{K}-band spectrum of the planet \citep{Rajan2017}. This was important in measuring accurate astrometry for the planet, because it avoided overfitting noise by down-weighting wavelengths where there was no contribution to the observed coherent fluxes from the planet's spectrum. The model of the coherent fluxes was first calculated on a 300x300 grid of RA and Dec across the 60 mas diameter field of view of the SC fiber, and then again on a 6 mas diameter grid centered on the fringe with the highest likelihood. The procedure is performed on each exposure throughout the observation, and the mean and covariance of the individual measurements are taken to be the final astrometry for the night.

\par The first GRAVITY observation of the system, taken in dual field off-axis mode in 2019-11-11, suffered from an uncorrectable metrology error that was incurred in-between the observation of the binary calibrator and the observation of the planet, and was discarded. An observation during poor observing conditions failed to detect the planet on 2020-02-02 and was discarded. Following the successful reconnaissance of the planet's position in the SPHERE observation detailed above, two successful observations in dual field off-axis mode were taken during ideal conditions on 2022-01-23 (Fig.~\ref{fig:detections}b) and  2022-09-15 (Fig.~\ref{fig:detections}c) and were retained. Observations on 2023-08-30, 2023-08-31, 2023-09-26, and 2023-11-25, without sufficient field rotation, resulted in non-detections and were discarded.

Observations taken in dual-field on-axis mode during the commissioning of the GRAVITY+ upgrade resulted in a detection on 2024-11-15 (Fig~\ref{fig:detections}d) that was off-center of the science camera fiber, due to a stale orbital solution used for the fiber pointing \citep{gravity_plus_firstlight}. Subsequent monitoring campaign observations taken during ideal conditions with sufficient field rotation based on an updated orbit fit resulted strong detections on 2024-12-18 (Fig~\ref{fig:detections}e) and 2025-10-03 (Fig~\ref{fig:detections}f). 

The relative astrometry of the planet is recorded in Table \ref{tab:astrometry}. The GRAVITY \textit{K}-band spectrum of the planet, published in \citet{Kammerer2025}, agrees within errors with the GPI \textit{K}-band spectrum from \citet{Rajan2017}.

\begin{deluxetable*}{ccccccc}
\tablewidth{\textwidth}
\tablecaption{New relative astrometry of 51~Eri~b. \label{tab:astrometry}}
\tablehead{
\colhead{Epoch [MJD]} & \colhead{$\Delta$RA [mas]} & \colhead{$\Delta$Dec [mas]} & \colhead{$\sigma_{\Delta\alpha}$ [mas]}  & \colhead{$\sigma_{\Delta\delta}$ [mas]} & \colhead{$\rho$} & \colhead{Instrument} \\
}
\startdata
59516.1      & 286.4  &   -231.9 &   3.7 &  4.2  & 0.0    &   SPHERE \\
59602.2      & 290.82 &  -218.89 & 0.17  &  0.37 &   -0.47 & GRAVITY \\
59837.3      & 294.55 &  -178.14 & 0.23  &  0.36 &   -0.3  & GRAVITY \\
60630.3      & 266.43 &  -24.53  & 0.24  &  0.26 &   -0.73 &   GRAVITY+ \\
60662.2 & 263.64 & -17.76 & 0.20 & 0.26 & -0.58 & GRAVITY+ \\
60952.3 & 226.13 & 42.46 & 0.18 & 0.24 & -0.51 & GRAVITY+ \\
\hline\hline
\enddata
\tablecomments{The co-variance matrix for each measurement can be reconstructed using $\sigma_{\Delta\alpha}^2$ and $\sigma_{\Delta\delta}^2$ on the diagonal, and $\rho\times\sigma_{\Delta\alpha}\times\sigma_{\Delta\delta}$ on the off-diagonal.}
\end{deluxetable*}

\subsection{JWST/NIRCam and Contrast-to-mass upper limits}

\par 51~Eri has been observed multiple times by JWST \citep{Rigby2023} in various coronagraphic modes, using both the wedge shaped coronagraphs to study 51~Eri~b in the near infrared \citep{Balmer2025b}, and the round and four-quadrant phase mask coronagraphs to search for additional, wide separation gas giants in the system. The latter observations have not yet been published. In this study, we briefly analyze one of these datasets in order to quantify the upper limits to additional planets within the system.

\par We considered the observations of the system with the Near Infrared Camera (NIRCam) \citep{Rieke2005, Rieke2023} \texttt{MASK430ROUND} coronagraph \citep{Krist2009, Girard2022} in the F444W filter ($\lambda_{\rm cen}=4.439\,\mathrm{\mu m}$, $\Delta\lambda_{\rm FWHM}=1.11\,\mathrm{\mu m}$), taken between UTC 2022-10-19 20:45:00 and 2022-10-20 07:18:30 (GTO 1412, PI: Perrin)\footnote{This dataset can be found on the Barbara A. Mikulski Archive for Space Telescopes (MAST) at \doi{10.17909/cgv3-9x97}.}. The observing sequence consisted of two observations of the target star at roll angles separated by $10^\circ$, and observations of a reference star (HD~30562) with a five point dither to increase the diversity of the point spread function library to mitigate random target acquisition errors \citep{Soummer2014}. The coronagraphic mask has an inner working angle (the point at which the coronagraphic transmission reaches 50\% attenuation) of 0\farcs82 (nearly $4\times$ the projected separation of 51~Eri~b at the time of observation), so 51~Eri~b is not intended to be visible in these observations.

\par We reduced, cleaned, and starlight subtracted the data using \texttt{spaceKLIP}\footnote{\url{https://spaceklip.readthedocs.io/en/latest/}}, a JWST high contrast imaging code suite that wraps the \texttt{jwst} data reduction pipeline \citep{Bushouse2025} for HgCdTe HAWAII-2RG detector data reduction and astronomical calibration\footnote{This work used pipeline and calibration file versions \texttt{jwst} v1.18; Calibration References Data System (CRDS) v12.1.5; CRDS context \texttt{jwst\_1364.pmap}.}, applies optional (but necessary for the best contrast performance) modifications for bad pixel cleaning, image registration, \citep{Kammerer2022, Carter2023, Balmer2025b}, and passes the data products to \texttt{pyKLIP}\footnote{\url{https://pyklip.readthedocs.io/en/latest/}} \citep{Wang2015} for starlight subtraction and source forward modeling with Karhunen--Lo\`{e}ve Image Projection (KLIP) \citep{Soummer2012, Pueyo2016, Wang2016}. Starlight subtraction was applied to the full subarray images without subdivision.

\par
\begin{figure}
    \centering
    \includegraphics[width=\linewidth]{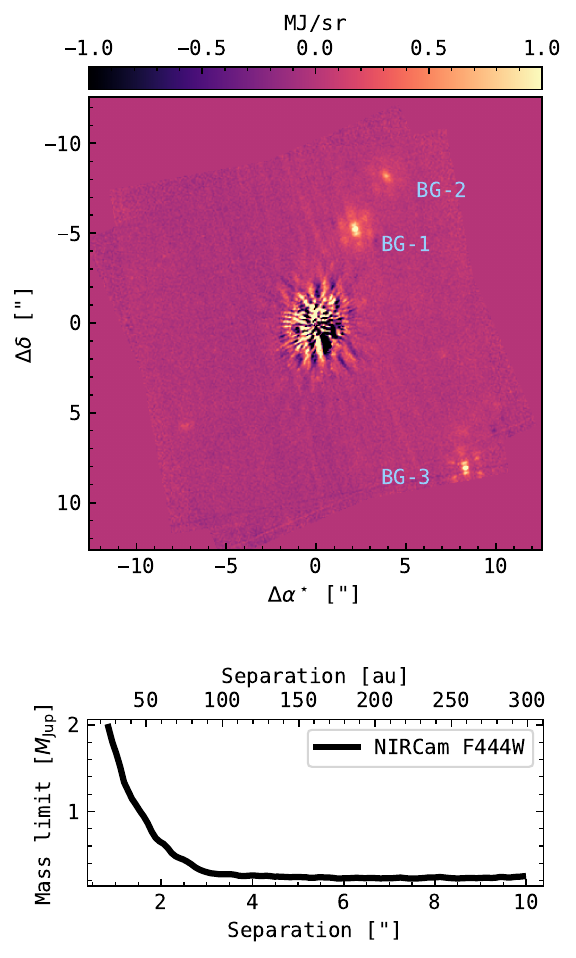}
    \caption{JWST coronagraphic imaging of 51~Eri and limits on additional planets. \textit{Top}: The starlight subtracted image in the F444W filter illustrates the detection of four background sources and the non-detection of any point sources consistent with planetary companions. \textit{Bottom}: Mass upper limits derived from the $5\,\sigma$ contrast curve in the F444W filter. These observations rule out additional planets exterior to 51~Eri~b down to Saturn masses past 100 au, and Jupiter masses past 25 au.}
    \label{fig:jwst_cc}
\end{figure}

 A number of resolved background sources are visible (galaxies), but no point sources are detected in the starlight subtracted image. We transformed the contrast into a mass upper limit using evolutionary models from \citet{Linder2019}, specifically interpolating their isochrones at the system's age \citep[$\sim24\,\mathrm{Myr}$,][]{Bell2015} and integrating the associated boundary condition \texttt{petitCODE} model atmospheres under the F444W filter\footnote{We used functionality in the \texttt{species} python package to do the interpolation of the evolutionary models to transform contrasts into mass upper limits \citep{Stolker2020}.}. We used the cloudiest ($f_{\rm sed}=1$), most metal rich ($\mathrm{[M/H]}=1.0\,\mathrm{dex}$) \texttt{petitCODE} model atmospheres, \added{as cloudier and more metal rich models suppress the continuum flux of a planet's spectrum, decreasing its brightness in the F444W filter. This choice therefore provide the largest (most conservative) mass upper limits. Additionally, we justify this choice based on recent observational evidence to suggest that cold and sub-Jovian mass planets have cloudy, metal rich atmospheres} \citep{BardalezGagliuffi2025, Bowens-Rubin2025, Crotts2025}. Figure \ref{fig:jwst_cc} illustrates the starlight subtracted F444W image and mass sensitivity curve. NIRCam achieves significantly deeper mass sensitivity ($<0.75\,\Mjup$ at 50 au and $\sim0.25\,\Mjup$ at 100 au) than the last published upper limits from deep SPHERE imaging \citep[$<2\,\Mjup$ at $>10$ au,][]{Maire2019}, ruling out much of the phase space where a third external body could induce secular oscillations on 51~Eri~b within the system's lifetime \citep{Denis2026}.

\section{Orbit Analysis}

\subsection{Orbit fitting}

Possible orbital solutions for the motion of 51~Eri~b were determined using \texttt{octofitter}\footnote{\url{https://sefffal.github.io/Octofitter.jl/dev/}} \citep{Thompson2023}, a comprehensive \texttt{julia} language \citep{Julia2017} orbit fitting code. The package allows the user to compare Keplerian orbits to a multitude of observations, including relative astrometry, relative radial velocities, absolute astrometry, and more, and estimate the posterior distribution on the orbital elements and other hyper-parameters with a number of samplers. 

Our fits considered the relative astrometric measurements of the planet from GPI \citep{DeRosa2020} and SPHERE \citep{Maire2019}, the relative (planetary) radial velocity from HiRISE \citep[][which were taken following predictions from preliminary orbits fit to our GRAVITY observations, see their Figure 2]{Denis2026}, and the absolute astrometric measurements of the host star, as recorded in the Hipparcos-Gaia Catalog of Accelerations \citep[HGCA,][DR3 verison]{Brandt2018, Brandt2021} and the Gaia DR2, DR3, Hipparcos Catalog \citep[G23H,][]{Thompson2026}. The G23H Catalog includes a measure of astrometric excess noise for the Gaia observations of each star called the unbiased estimator of variance \citep[UEVA,][]{Kiefer2025}, estimated from the Renormalized Unit Weight Error (RUWE) for the host star in the Gaia DR3 catalog. For completeness, we retabulate the HGCA and G23H entries for 51 Eridani in Table \ref{tab:absastrometry}. Sampling was conducted with \texttt{Pigeons.jl} \citep{Surjanovic2023}, a non-reversible parallel tempered MCMC algorithm. We ran the sampler with 64 chains, and for 17 rounds (where adaptation is performed in-between rounds, and the number of samples are subsequently doubled). 

\begin{deluxetable}{ccCCC}
\tablewidth{\textwidth}
\tablecaption{Absolute astrometry for 51~Eri considered in this work. \label{tab:absastrometry}}
\tablehead{
\colhead{Quantity} & \colhead{Source} & \colhead{Epoch} &\colhead{Value} & \colhead{Uncertainty}
}
\startdata
\multicolumn{5}{c}{In HGCA} \\
$\mu_{\alpha^*}$ & HIP & 48225.7  &  43.6733   & 0.609772  \\
$\mu_\delta$     & HIP & 48156.6  &  -64.1988  & 0.475111  \\
$\rho_{\mu_\alpha^*,\mu_\delta}$ & HIP. & \nodata & 0.0140913 & \nodata \\
$\mu_{\alpha^*}$ & H-G3 & 52763.2  &  44.2575   & 0.0186419 \\
$\mu_\delta$     & H-G3 & 52746.9  &  -64.1774  & 0.0109409 \\
$\rho_{\mu_\alpha^*,\mu_\delta}$ & H-G3. & \nodata & -0.126630 & \nodata \\
$\mu_{\alpha^*}$ & G3   & 57300.6  &  44.0488   & 0.138176  \\
$\mu_\delta$     & G3   & 57337.3  &  -64.0284  & 0.102076  \\
$\rho_{\mu_\alpha^*,\mu_\delta}$ & G3 & \nodata & -0.453427 & \nodata \\
\hline
\multicolumn{5}{c}{In G23H} \\
$\mu_{\alpha^*}$ & HIP & 48225.7 & 43.6733   & 0.609772  \\  
$\mu_\delta$     & HIP & 48156.6 & -64.1988  & 0.475111  \\ 
$\rho_{\mu_\alpha^*,\mu_\delta}$ & HIP & \nodata & 0.014091 & \nodata \\
$\mu_{\alpha^*}$ & H-G3 & 52763.2 & 44.2575   & 0.0186419 \\   
$\mu_\delta$     & H-G3 & 52746.9 & -64.1774  & 0.0109409 \\
$\rho_{\mu_\alpha^*,\mu_\delta}$ & H-G3 & \nodata & -0.126630 & \nodata \\
$\mu_{\alpha^*}$ & G2   & 57254.5 & 44.403    & 0.621834  \\ 
$\mu_\delta$     & G2   & 57288.5 & -63.7168  & 0.487668  \\
$\rho_{\mu_\alpha^*,\mu_\delta}$ & H-G3 & \nodata & -0.432174 & \nodata \\
$\mu_{\alpha^*}$ & G2-3 & 57277.6 & 42.4448   & 1.10365   \\
$\mu_\delta$     & G2-3 & 57312.9 & -64.6005  & 0.758062  \\
$\rho_{\mu_\alpha^*,\mu_\delta}$ & G2-3 & \nodata & -0.489372 & \nodata \\
$\mu_{\alpha^*}$ & G3   & 57300.6 & 44.0488   & 0.138176  \\ 
$\mu_\delta$     & G3   & 57337.3 & -64.0284  & 0.102076  \\ 
$\rho_{\mu_\alpha^*,\mu_\delta}$ & G3 & \nodata & -0.453427 & \nodata \\
RUWE             & G3   & 57312.9 & 1.14529   & \nodata       \\ 
\enddata
\tablecomments{Epochs are the mean across the observation, though the observing baseline for each measurement is accounted for in the orbit fits as described in \citet{Thompson2026}. HGCA values are from the EDR3 catalog in \citet{Brandt2021}. G23H values are from the catalog in \citet{Thompson2026}.}
\end{deluxetable}

\par Our fiducial fit was a one planet (two body) Keplerian orbit fit to the relative measurements made of 51~Eri~b. We then considered the same fit including additionally the absolute astrometry of the host star from the HGCA, and then a third fit considering the G23H absolute astrometry. 

\begin{figure}
    \centering
    \includegraphics[width=\linewidth]{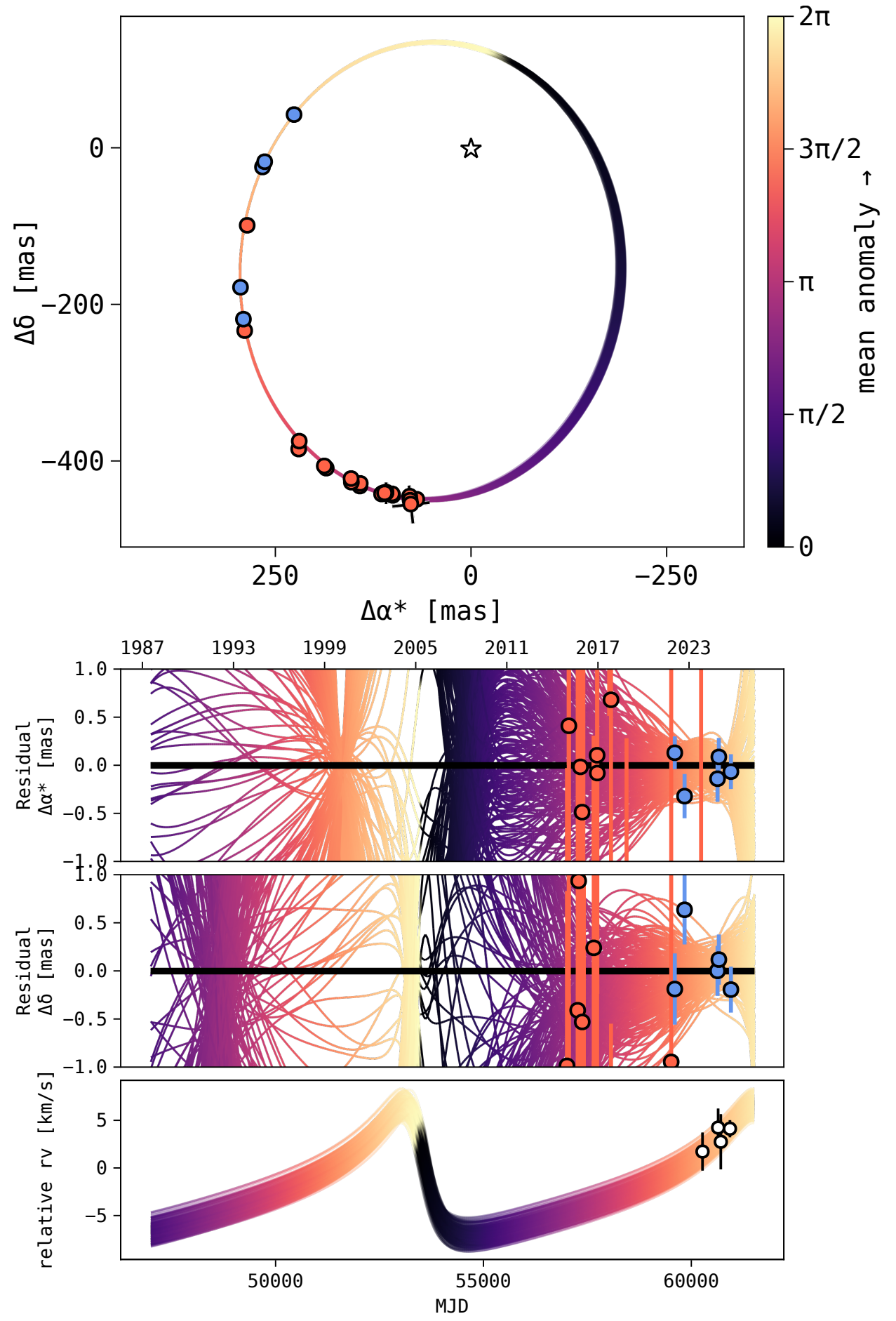}
    \caption{The results of the fiducial orbit fit including all relative astrometry and relative radial velocities. \textit{Top}: Visual orbit of the planet; measurements from GRAVITY are plotted as red scatter points, from coronagraphic telescopes (Gemini, Keck, VLT, JWST) as blue scatter points. 250 random orbits from the posterior distribution are color-coded according to their mean anomaly $M$, the fraction of the orbital period that the planet has completed since its closest approach to the star. \textit{Middle}: Residuals from the maximum a-posteriori orbit in R.A. and Decl. versus time, for the same orbits and observations. \textit{Bottom}: Relative radial velocity versus time; HiRISE observations are plotted as white scatter points.}
    \label{fig:fiducial_orbit}
\end{figure}

The posterior distribution on the parameters of the fiducial (relative measurements of planet b only) model fit are recorded in Table \ref{tab:orbit}. This orbit fit is visualized in Figure \ref{fig:fiducial_orbit}; as shown in that figure, there are residuals between the high precision GRAVITY relative astrometry and the maximum a-posterior orbit on the order of 0.5 mas. 

In all cases, when fitting orbits to the planetary relative radial velocities from HiRISE \citep[][their Table 1]{Denis2026}, best fit models included a zero-point offset term that converged to $-2.3\pm0.7\,\mathrm{km/s}$. This appears due to the uncertainty in the stellar radial velocity they assumed from the Pulkovo Compilation of Radial Velocities \citep[$v_\star=12.6\pm0.3\,\mathrm{km/s}$,][]{Gontcharov2006}. This radial velocity is actually a combination of the results of four publications, with a standard deviation of $7.0\,\mathrm{km/s}$\footnote{\url{https://vizier.cds.unistra.fr/viz-bin/VizieR?-source=III/252}, Table 8}. Our fitted offset implies $v_\star=10.3\pm1.0\,\mathrm{km/s}$, well within the dispersion of the Pulkovo values that underpin the assumption in \citet{Denis2026}.  

\par
\begin{figure}
    \centering
    \includegraphics[width=\linewidth]{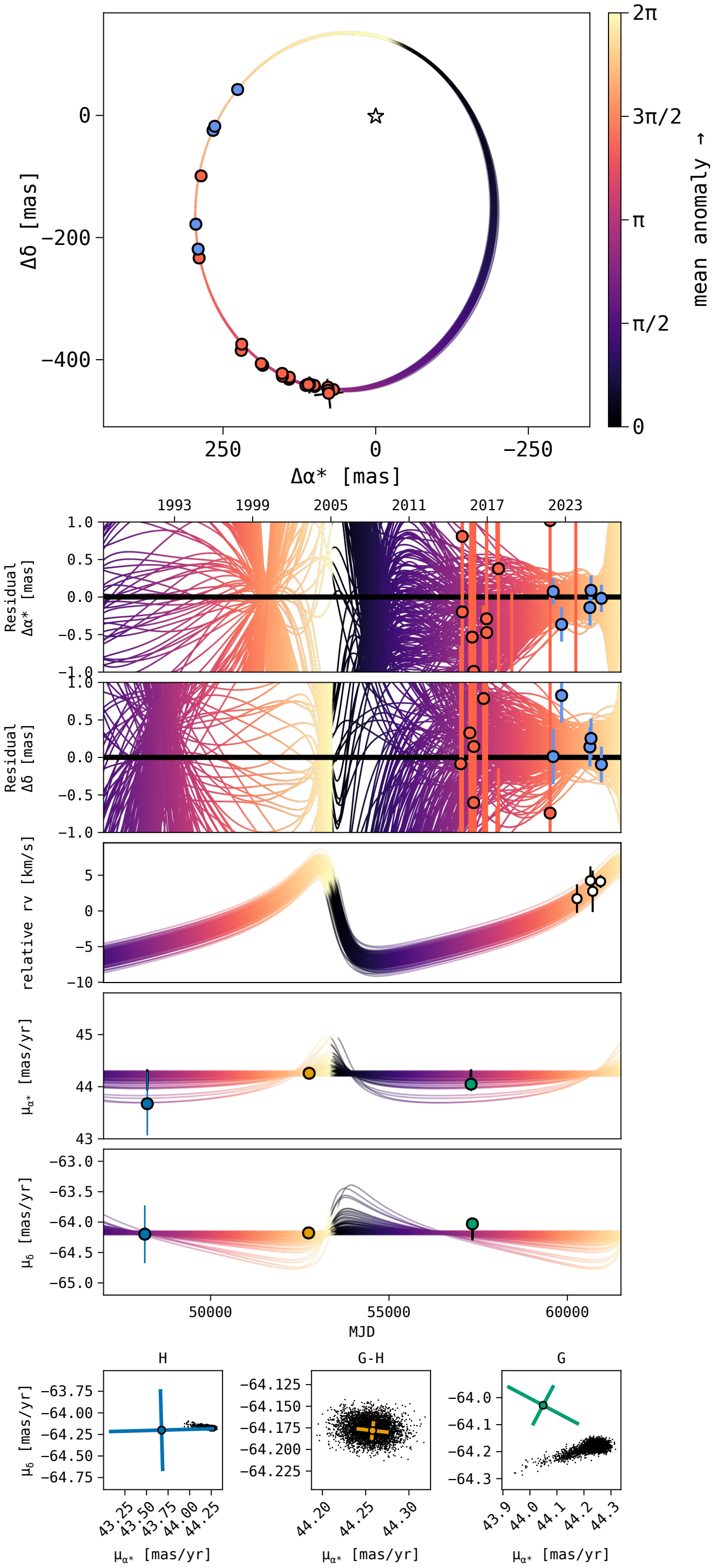}
    \caption{The results of the orbit fit adding the HGCA. \textit{Top four panels}: As in \ref{fig:fiducial_orbit}. \textit{Next two panels}: Proper motion in R.A. and Decl. of the system over versus time; the Hipparcos, Hip.-Gaia, and Gaia DR3 proper motions are plotted as scatter points. \textit{Bottom row}: Proper motions in R.A. versus Decl. (including their correlation, shown as the rotation of the errorbars) for each epoch of observation in the HGCA, illustrating the tension between the motion predicted by the single planet orbit and the Gaia DR3 astrometry.}
    \label{fig:hgca_1p_orbit}
\end{figure}

 Although the planet was near pericenter during the Hipparcos observation, the planet is too low mass and the system too distant for the planet's reflex motion to have influenced the Hipparcos intermediate astrometric observations appreciably. The HGCA is in principle more sensitive to the period and amplitude of the orbit, but does not exhibit acceleration consistent with the observed planet's orbit. The Gaia DR3 proper motion (especially in Decl.) recorded in the HGCA is in tension with the anticipated reflex motion of 51~Eri~b's orbit, as can be seen in Figure \ref{fig:hgca_1p_orbit}. This has been interpreted as effectively setting an upper limit on the 51~Eri~b's mass \citep{DeRosa2020, Dupuy2022}. Fitting one planet orbits to relative astrometry, relative radial velocities, and the HGCA, we find this interpretation to be prior dependent, varying as to the choice of a log-uniform prior on mass ($<12\,M_{\rm J}$ at $3\,\sigma$) or uniform prior ($<15\,M_{\rm J}$ at $3\,\sigma$), as is commonly seen for low signal-to-noise accelerations in the HGCA \citep[e.g.][]{Bovie2025}. We summarize the visual orbital elements for the one planet HGCA orbit fit in Table \ref{tab:orbit}.

\begin{figure*}
    \centering
    \begin{subfigure}[c]{0.48\textwidth}
    \includegraphics[height=0.7\paperheight]{cEri-1p-g23h-ueva-mcmc-orbits.pdf} 
    \caption{}
    \end{subfigure}
    \hfill
    \begin{subfigure}[c]{0.48\textwidth}
    \includegraphics[height=0.7\paperheight]{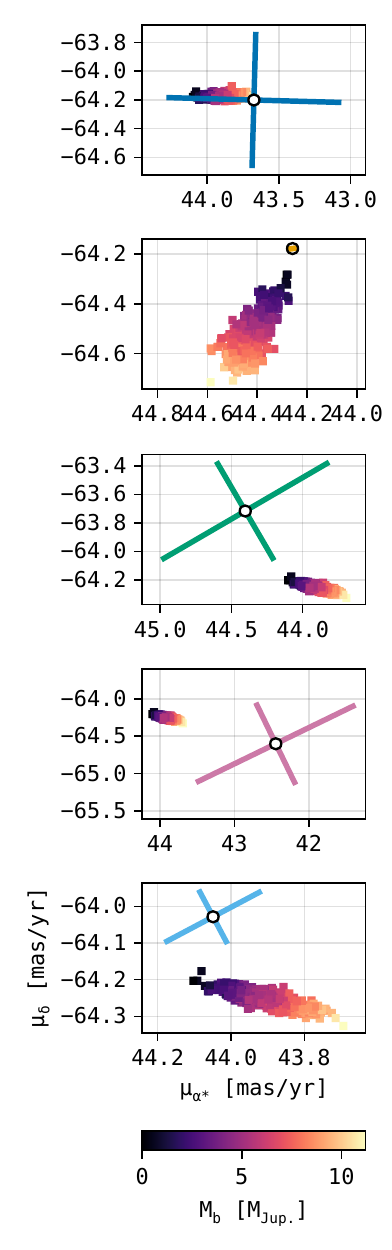}
    \caption{}
    \label{fig:subim2}
    \end{subfigure}
    \caption{The results of the orbit fit including the G23H absolute astrometry. (a) \textit{Top and middle panels}: same as in Figure \ref{fig:hgca_1p_orbit}. \textit{Bottom}: Z-scores (normalized residuals, equivalent to $\sigma$ standard deviations) for each absolute astrometric measurement from the G23H catalog. A black horizontal line denotes a z-score of 0, and gray horizontal lines mark z-scores of 1. A larger z-score indicates that the measurement is in greater tension with the model. (b) \textit{Top to bottom}: The G23H catalog values for the proper motion of the system (colored errorbars, which include the correlations between R.A. and Decl.), compared against the instantaneous predictions from the best fit linear motion and one planet orbit (colored squares, shaded by the mass of the planet). This visualization provides a more detailed view of the data and goodness of fit metric in the lower half of (a).}
    \label{fig:g23h_1p_orbit}
\end{figure*}

The time-baseline covered by the Gaia DR2 and DR3 absolute astrometry is small compared to the orbital period, and the Gaia observations were taken when the planet was near apocenter. Considering the G23H astrometry in a single planet fit to the orbit of b results in a suspiciously precise dynamical mass for 51~Eri~b of $M_{\rm b}=6.8\pm1.4\,M_{\rm J}$. As Figure \ref{fig:g23h_1p_orbit}b shows, this appears driven by the tension between the Hip-Gaia, and Gaia DR2 and Gaia DR3 proper motions (which prefer solutions with negligible mass for b's orbit) and the DR2-DR3 proper motion (which prefers higher mass solutions). In any case, neither massive ($M_{\rm b}\sim7\,M_{\rm J}$) or massless orbits can fully reproduce the proper motion measurements. We therefore caution against interpreting this result as evidence for any dynamical mass constraint on b, and more generally against using short-term proper motion anomalies to fit the orbits of companions with significantly longer periods than the absolute astrometric baseline.

\begin{figure*}
    \centering
    \includegraphics[width=0.85\linewidth]{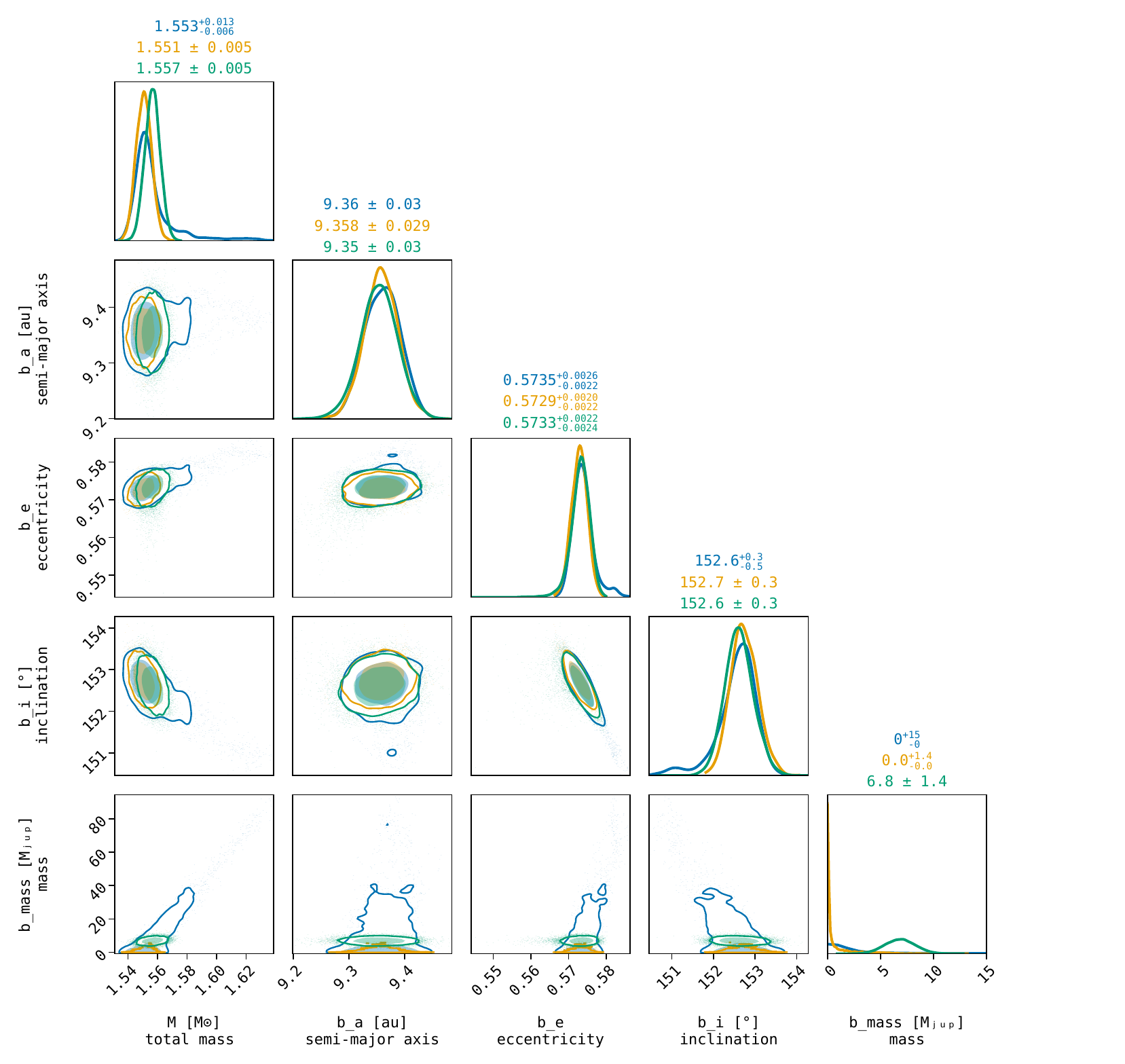}
    \caption{Comparison between the posterior distributions on key orbital parameters for orbit fits considering different data (fiducial, relative astrometry and relative RV only: blue, plus HGCA: orange, plus G23H: green). The fiducial dataset and the HGCA place upper limits on the dynamical mass of 51~Eri~b. The G23H fit ``constrains" the dynamical mass, but appears to suffer from overfitting to the absolute astrometry, based on the results of our leave-one-out tests. Our new high precision GRAVITY+ relative astrometry constrains the eccentricity of the planet's orbit to $e=0.573\pm0.003$, regardless of choice of absolute astrometric dataset.}
    \label{fig:small_corner_compare}
\end{figure*}

Although it seems highly unlikely to be driven by the known planet, it is possible that the proper motion anomalies observed between Hipparcos, Gaia DR2, and Gaia DR3 are caused by an additional, inner companion. We therefore considered two planet (three body) orbits fit to this ensemble of data. This was motivated by the additional constraint that the the reflex motion induced on the star by the putative inner companion would be constrained not only by the absolute astrometry, but by our high precision relative astrometry from GRAVITY, which is a measure of the planet-star distance, not the planet-barycenter distance \citep{Lacour2021}. We tested two scenarios, 1) the inclination of the two companions are independent, $P(i_c)=\sin(i_c)$, $P(\Omega_c)=\mathcal{U}(0,2\pi)$, 2) the two companions are coplanar, so we sample only one set of orbital inclination $i_{\rm planetary}$ and longitude of ascending node $\Omega_{\rm planetary}$. We motivate the study of the coplanar case by noting that simulations of two-planet planet-planet scattering achieve high eccentricities, but rarely high mutual inclinations \citep[e.g.][]{Gratia2017}. \added{Due to the prior dependence of the mass estimate in the 1-planet HGCA fit, we then investigated the choice of the log-uniform prior on the masses of the two companions. This choice of prior for the fiducial case was motivated by the default prior adopted in \texttt{orvara} \citep{Brandt2021b} and therefore previous work on the dynamical mass limits of b \citep{Dupuy2022}, but may be inappropriate for these potentially massive companions and incomplete absolute astrometry, as it could weight the posterior towards solutions with vanishing mass.}

\begin{deluxetable*}{ccCCCCCCC}
\tablewidth{\textwidth}
\footnotesize
\tablecaption{Orbital parameters inferred for 51~Eri~b and ``c" in this work. \label{tab:orbit}}
\tablehead{
\colhead{Parameter} & \colhead{Prior} & \multicolumn{7}{c}{Posterior} \\
\colhead{} & \colhead{} & \colhead{b, 1p rel.} & \colhead{b, 1p +HGCA} & \colhead{b, 1p +G23H} & \colhead{b, 2pLN} & \colhead{c, 2pLN} & \colhead{b, 2pU200} & \colhead{c, 2pU200}
}
\startdata
$a$ [au]                              & $\log\,\mathcal{U}(0.001, 10^4)$   & 9.36\pm0.03        & 9.36\pm0.03         & 9.35^{+0.04}_{-0.03} & 9.43^{+0.09}_{-0.16} & 0.3-3.14 & 9.36\pm0.08 & 0.1-2.4 \\
$e$                                   & $\mathcal{U}(0,1)$                & 0.574\pm0.003      & 0.573\pm0.003       & 0.573\pm0.003      & 0.57^{+0.01}_{0.01} & 0.0-0.8 & 0.58^{+0.01}_{-0.01} & 0.0-1.0 \\
$i$ [$^\circ$]                        & $\mathrm{Sine}(0,90)$            & 152.6^{+0.5}_{-0.2} & 152.7\pm0.3         & 152.6^{+0.3}_{0.4}        & 151.8_{-0.6}^{+0.6} & 36-174 & 151.8^{+1.1}_{-1.2} & 27.0-169.0 \\
$\omega$ [$^\circ$]                 & $\mathcal{U}(0,180)$               & 58.4^{+1.2}_{-1.1}  & 58.3_{-1.1}^{+1.2}  & 57.88^{+1.58}_{1.04} & 60.2_{-9.2}^{+4.6} & -168-+173 & 57.3^{+3.6}_{-3.7} & -177-+164 \\
$\Omega$ [$^\circ$]                 & $\mathcal{U}(0,180)$               & 37.1^{+2.0}_{-1.2}  & 37.2_{-1.6}^{+1.4}  & 36.56^{+2.02}_{1.46} & 39.5_{-12.6}^{+5.7} & -163-+180 & 35.1^{+4.7}_{-5.0} & -178-+167 \\
$\theta$ [$^\circ$]                       & $\mathcal{U}(0,180)$         &142.2^{+0.04}_{-0.03}& 142.20\pm0.03       & 116.88^{+0.02}_{0.03}      & 116.8_{-0.06}^{+0.06} & -180-+157  & 116.9^{+0.1}_{-0.1} & -180.0-+162.0 \\
$\pi$ [mas]                         & $\mathcal{N}(33.439, 0.077)$        & 33.45\pm0.07       & 33.44\pm0.07        & 33.44\pm0.08       & 33.4^{+0.09}_{0.11} & \nodata & 33.44^{+0.15}_{-0.15} & \nodata \\
$M_\mathrm{p}$ [$M_\mathrm{J}$]     & $\frac{\log\,\mathcal{U}(0.01,1000)}{\mathcal{U}(2,200)}^\dag$      & 0^{+15}_{-0}  & 0.0_{-0.0}^{+1.4}   & 6.8^{+1.4}_{-1.5}  & 4.5^{+1.8}_{-4.5} & 1.0-2900 & 10.9^{+8.7}_{-8.9} & 2.0-72.81 \\
$M_\mathrm{A}$ [$M_\odot$]          & $\mathcal{N}(1.550,\,0.006)$        & 1.550\pm0.005      & 1.551\pm0.005       & 1.550\pm0.005      & 1.571^{+0.008}_{0.014} & \nodata & 1.59^{+0.04}_{-0.03} & \nodata \\
$\ln{\mathcal{L}}$ & \nodata & -147.7^{+3.0}_{-1.5} & -141.9^{+2.6}_{-1.8} & -361.3^{+3.8}_{-2.6} & -358.7^{+3.9}_{-2.8} & -348.8^{+5.0}_{-3.3} & -358.7^{+3.0}_{-2.6} & -347.9^{+1.4}_{-0.9} \\
$\ln{\mathcal{P}}$ & \nodata & -157.6^{+3.6}_{-1.7} & -154.3^{+3.2}_{-1.8} & -467.1^{+7.0}_{-5.9} & -480.3^{+7.4}_{-5.7} & -469.9^{+8.0}_{-6.3} & -480.7^{+2.3}_{-1.6} & -469.9^{+2.0}_{-1.0} 
\enddata
\tablecomments{We report the mean and 68\% confidence interval on each parameter for the orbit of 51~Eri~b, and the 95\% confidence interval on each parameter for the orbit of 51~Eri~``c." $\theta$ is the position angle of the orbit at the reference epoch, 2020-01-01. \added{$\ln{\mathcal{L}}$ and $\ln{\mathcal{P}}$ represent the log-likelihood and log-posterior probability, respectively. Importantly, the values under the b, 2-planet and c, 2-planet column represent the probability for the chains where only b is included, and where b and c are included. \dag The default prior on the companion masses is the wide log uniform prior, but we tested this prior explicitly in the two companion case by assuming a uniform prior between $2-200\,M_{\rm J}$ instead.}}
\end{deluxetable*}

\begin{figure*}
    \centering
    \begin{subfigure}[c]{0.6\textwidth}
    \includegraphics[height=0.6\paperheight]{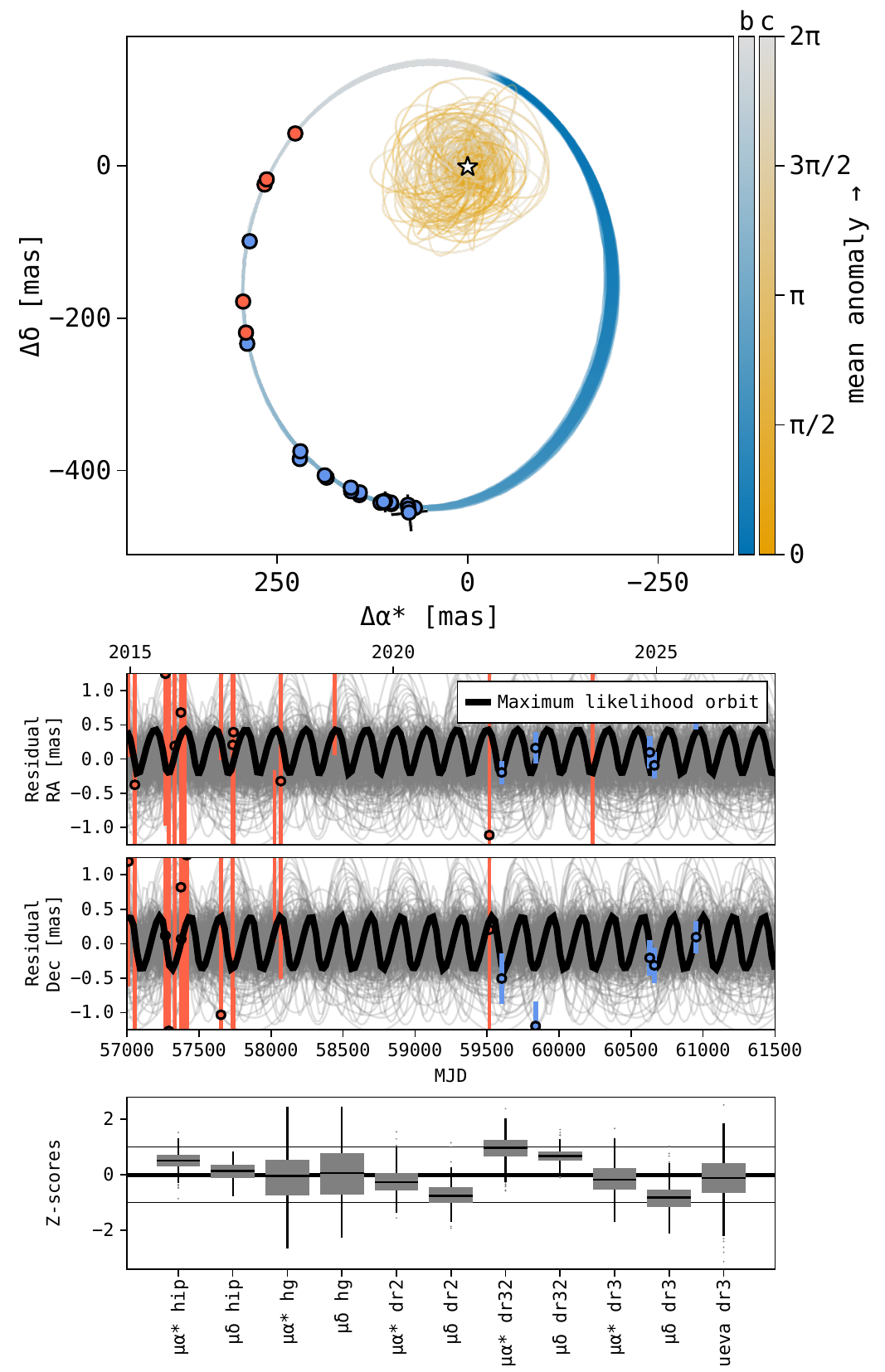} 
    \caption{}
    \end{subfigure}
    \hfill
    \begin{subfigure}[c]{0.35\textwidth}
    \includegraphics[height=0.6\paperheight]{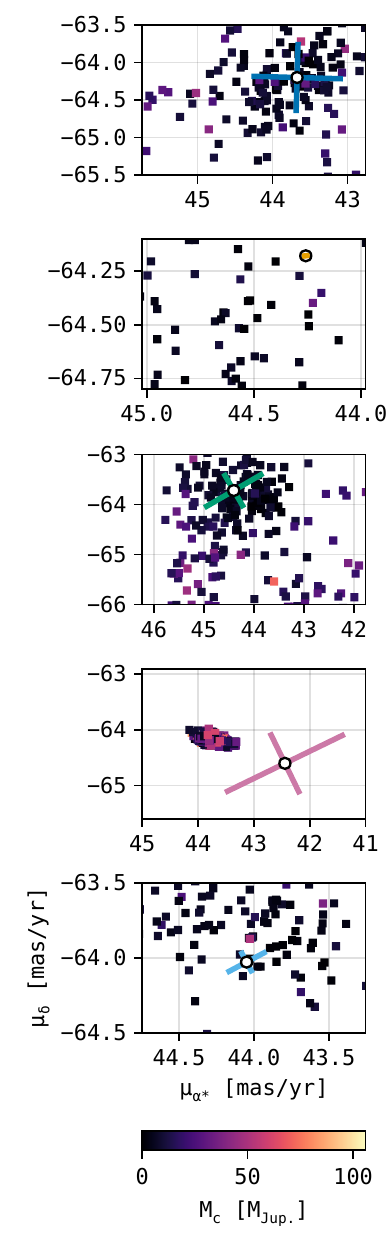}
    \caption{}
    \label{fig:subim2}
    \end{subfigure}
    \caption{Overview of the free inclination two-planet orbit fit.  (a) Visual orbits (top), Residuals after subtracting the orbit of b in R.A. and Decl. (middle), and astrometric z-scores (bottom) are plotted as in Figure \ref{fig:g23h_1p_orbit}. 
    The visual orbits for both the inner companion (orange) and known planet (blue) are shown. The residual motion of 51~Eri~b due to 51~Eri~c's influence on the host star is shown in the middle panels in R.A. (top middle) and Decl. (bottom middle). The maximum likelihood solution for the orbit of 51~Eri~b from the two planet, coplanar orbit fit has been subtracted from the observations and the orbit samples. In contrast to the one planet model, the addition of a second companion brings the z-scores for this model to within 1 for all the astrometric measurements. (b) Proper motions as in Figure \ref{fig:g23h_1p_orbit}, except that the model now includes the candidate ``c" and the model squares are shaded by the mass of ``c".}
    \label{fig:orbit_2p_overview}
\end{figure*}

\begin{figure*}
    \centering
     \includegraphics[width=\linewidth]{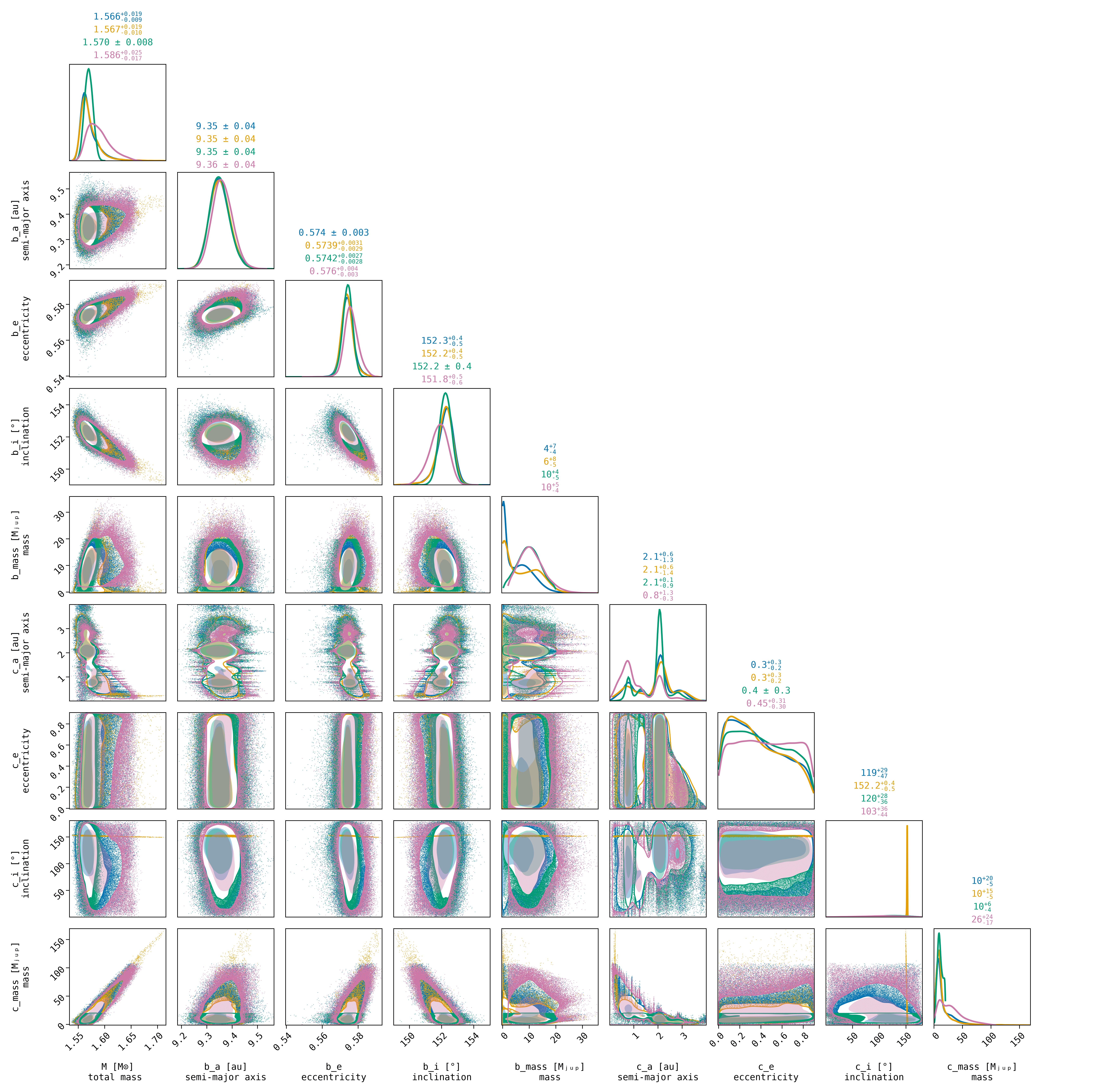}
    \caption{Comparison between the posterior distributions on key orbital parameters for the two-planet orbit fits. Isotropic mutual inclinations, log uniform mass prior: blue, co-planar mutual inclinations, log uniform mass prior: orange, isotropic $i$, uniform mass prior between $0.2-20\,M_{\rm J}$: green, isotropic $i$, uniform mass prior between $2-200\,M_{\rm J}$: pink. The orbital elements for b are largely unchanged from the one-planet orbit fits. The current observations constrain the semi-major axis of the putative inner companion to peaks in the probability distribution at 2.1 and 0.8 au, and limit the mass of this companion to be $\lesssim20\,\Mjup$ at $1\sigma$. The coplanar constraint enhances the probability of the $2.1\,\mathrm{au}$ mode of c's semi-major axis, though our stability analysis indicates the majority of the orbits in this mode are unstable (see \S\ref{subsec:dynstab}. The correlation between the mass of b and ``c" reveals the reason for the apparent peak in the mass of b in the one-planet orbit fit including the G23H astrometry in Figure \ref{fig:small_corner_compare}: the excess proper motion anomaly in the Gaia data can be explained by an inner companion rather than the known planet.}
    \label{fig:2p_corner}
\end{figure*}

In each two-companion orbit fit, we placed an additional requirement that the orbits of the two companions do not cross. Figure \ref{fig:orbit_2p_overview} illustrates the results of the free-inclination fit, and Figure \ref{fig:2p_corner} shows the marginalized posterior distribution on the key orbital elements of the two companions from both the coplanar, free inclination, and Uniform[0.2,20] and Uniform[2,200] mass prior fits. We summarize the visual orbital elements for the free inclination, log-uniform mass prior, two planet fit in Table \ref{tab:orbit}; in this table we only record the 95\% confidence interval on the parameters associated with the second planet c (and not the central values), e.g. $M_c\in[1.0, 2900]_{95\%}\,M_{\rm Jup}$ and $a_c=[0.3, 3.1]_{95\%}\,\mathrm{au}$.

The inner companion's mass and semi-major axis are strongly correlated, with brown dwarf and stellar masses ($M_c\gtrsim13\,\Mjup$) restricted to shorter orbital separations ($a_c\lesssim1\,\mathrm{au}$) and the more likely, planetary mass ($1{-}13\,\Mjup$) samples centered on $a_c=2.1\,\mathrm{au}$. The eccentricity of the inner companion is unconstrained by the available data, as are the orbital angles, which reproduce their priors. Due to the degeneracy between $M_c$ and $a_c$, the choice of prior on the masses of the two components modulates the posterior distribution of periods for the putative inner companion. Figure \ref{fig:2p_corner} illustrates this, showing how the broader distribution on $M_c$ from the uniform mass prior fits reuslts in more posterior weight in the smaller period (lower $a_c$) modes of the distribution, particularly down-weighting the $a_c\sim2.1\,\mathrm{au}$ mode in favor of the $a_c\sim0.8\,\mathrm{au}$ mode.

\added{Following \citet{Blakely2026}, we implemented a discrete ``indicator variable" in our multi-planet orbit fits called $c_{\rm exists}$, which has a Bernoulli distribution prior centered on 0.5 (equal odds of returning 0 or 1). For each sample, this parameter is multiplied against the mass of the putative companion. \texttt{Pigeons.jl} allows for mixed sampling of discrete and continuous parameters, enabling a dynamic hypothesis test within the sampling process. If $c_{\rm exists}$ is 1, the mass for the putative companion is equal to the value drawn from the prior on the mass, but if $c_{\rm exists}$ is 0, the mass of the third body vanishes, leaving a sample that is equivalent to a two-body orbit between only 51~Eri~A and b. This is numerically equivalent to a trans-dimensional model that varies the number of parameters draw to draw, as in some analyses of radial velocity exoplanet detections \citep{Brewer2015}. With this indicator variable, one can compute a Bayes factor on the hypothesis test from within the posterior distribution resulting from a single sampling run. The Bayes factor for this hypothesis test is defined as ${\rm BF}=\frac{\rm P(data|N_p=2)}{\rm P(data|N_p=1)}=\frac{\bar{c}_{\rm exists}}{1-\bar{c}_{\rm exists}}$, where $\bar{c}_{\rm exists}$ is the mean of the indicator variable. This is because, in using the Bernoulli prior centered on 0.5, the prior odds are 1:1 and therefore the posterior odds are the Bayes factor \citep{MOREY2016}.}

We examined the Bayes factor for this hypothesis test on the presence of ``c". In all cases, (co-planar and free inclination with a log-uniform mass prior, and both runs with varied uniform mass priors), the sampler found strong evidence for the inclusion of a third body given the data. After 17 rounds of sampling, the resulting Bayes Factor (BF) was found to be $\rm BF=127.44^{+7.31}_{-6.8}$, or $\sim130$:1 odds in favor of the multi-planet interpretation for the free inclination, log-uniform mass prior case. The exact posterior distribution on this Bayes factor is sensitive to the prior on the masses of the two companions, as is the physical nature of the system architecture (as discussed later), but the overall evidence in favor of a third body is not sensitive to this choice. For the free inclination, uniform mass prior between $0.2,20\,M_{\rm J}$, we find $\rm BF=329.63^{+28.12}_{-25.25}$, and for the uniform prior between $2,200\,M_{\rm J}$, we find $\rm BF=96.52^{+6.78}_{-6.2}$. So, restricting the masses of both bodies to the planetary mass regime enhances the odds in favor of the three-body orbit fit, while restricting the masses to uniform values spanning super-Jupiters to low mass stars only marginally decreases the evidence. Regardless of the choice of mass prior, the odds in favor of the three-body orbit fit is still strong (order $10^2$).

The reason the sampler favors the inclusion of a second companion becomes obvious when examining the residuals between the data and the two planet model; the residuals to the absolute astrometry were brought within $1\,\sigma$ in $\mu_{\alpha^\star}$ and $\mu_{\delta}$, compared to $2-3\,\sigma$ in the one planet fit, and the relative astrometry of the outer planet can be marginally explained given the added oscillations on the outer planet-star distance. \added{This is also shown in the likelihoods $\mathcal{L}$ between the single planet G23H and the two planet free inclination orbit fits (equivalent to the hypothesis test with the internal indicator variable). Table \ref{tab:orbit} shows that the single planet fit gives $\ln{\mathcal{L}}=-361.3^{+3.8}_{-2.6}$, the fraction of the posterior with only one planet in the free inclination posterior gives $\ln{\mathcal{L}}=-358.7^{+3.9}_{-2.8}$, and the fraction with two planets gives $\ln{\mathcal{L}}=-348.8^{+5.0}_{-3.3}$.} In this orbit fit, the motion of b also exhibits periodic deviations from a two-body Keplerian orbit that could be caused by the reflex motion induced on the star by the inner companion. 

\subsection{Dynamical Stability} \label{subsec:dynstab}
To investigate the dynamical viability of a second companion in the 51~Eri system, we assess the stability of the orbital configurations drawn from the joint posterior distribution of 51~Eri~b and the candidate companion ``c''. For each posterior sample in the free inclination, log uniform mass prior orbit fit, we integrate the corresponding three-body system over \(T = 2~\mathrm{Myr}\), using a time step of \(0.0025~\mathrm{yr}\), with the symplectic integrator SABAC4 \citep{2001CeMDA..80...39L}. The stability of each solution is then evaluated through frequency analysis \citep{1990Icar...88..266L, 1993PhyD...67..257L}. Following \citet{2005A&A...440..751C, 2010A&A...511A..21C}, we compute the main frequency from the mean longitude of the outer planet over two consecutive time intervals, \([0,T/2]\) and \([T/2,T]\). We define the stability index as \(D = \left|1 - n^\prime/ n\right|\),
where \(n\) and \(n^\prime\) are the main frequencies measured over the first and second time intervals, respectively. Values of \(D\) close to zero indicate regular solutions, whereas larger values are associated with strongly chaotic motion \citep{1993PhyD...67..257L}.

\begin{figure}
	\centering
	\includegraphics[width=\linewidth]{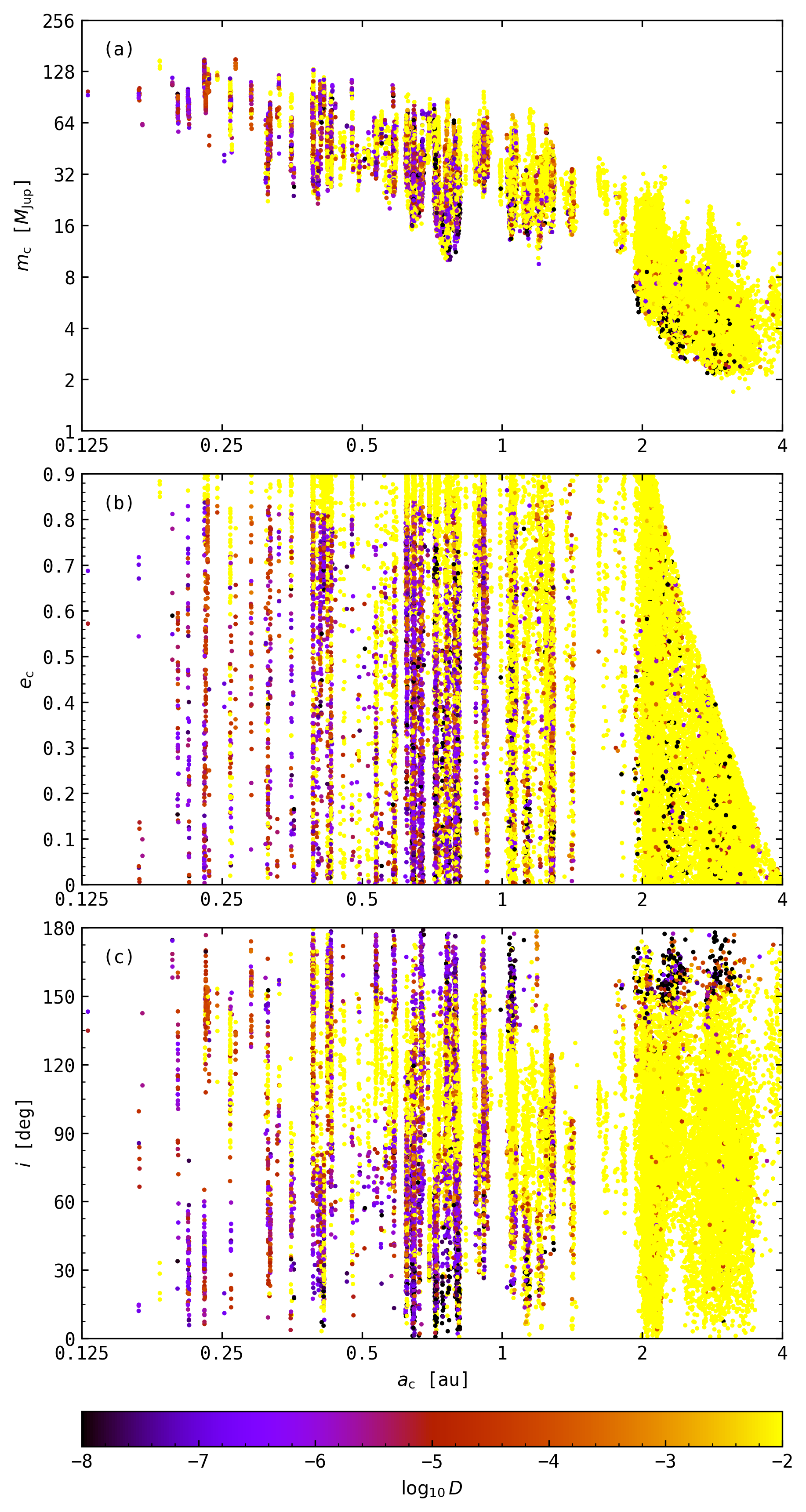}
	\caption{
    Stability analysis of the 51~Eri system with a candidate second companion, applied directly to the joint posterior distribution. Each point corresponds to one posterior sample for which the system was integrated and assigned a stability index, \(D\). The top panel shows the posterior samples in the semi-major axis--mass plane of the candidate companion, \((a_{\mathrm{c}},m_{\mathrm{c}})\), the middle panel shows the corresponding semi-major axis--eccentricity plane, \((a_{\mathrm{c}},e_{\mathrm{c}})\), and the bottom panel shows the corresponding semi-major axis--mutual-inclination plane, \((a_{\mathrm{c}},i)\). The panels are colour-coded by \(\log_{10} D\), with dark purple colours indicating dynamically regular configurations and yellow colours indicating strongly chaotic solutions.}
	\label{fig:Eri51_b_freq_analysis}
\end{figure}

In Fig.~\ref{fig:Eri51_b_freq_analysis}, we show the posterior samples in the \((a_{\mathrm{c}},m_{\mathrm{c}})\) plane (top), in the \((a_{\mathrm{c}},e_{\mathrm{c}})\) plane (middle), and in the \((a_{\mathrm{c}},i)\) plane (bottom), colour-coded by the stability index \(\log_{10}D\). Here, \(i\) represents the mutual inclination between 51~Eri~b and the candidate companion ``c'' \citep[e.g.][]{2012A&A...541A.151G},
\begin{equation}
    \cos i
    =
    \cos I_{\rm c}\cos I_{\rm b}
    +
    \sin I_{\rm c}\sin I_{\rm b}
    \cos\left(\Omega_{\rm c}-\Omega_{\rm b}\right),
\end{equation}
where \(I_k\) and \(\Omega_k\) are the inclination and longitude of the ascending node, respectively, with \(k=b,c\).

Colors close to ``yellow'' indicate chaotic motion, whereas colors close to ``dark purple'' indicate regular solutions. The frequency analysis shows that the dynamically regular configurations occupy a restricted region of the posterior distribution. In particular, a significant fraction of low-diffusion solutions is concentrated at semi-major axes \(a_{\mathrm{c}} < 1~\mathrm{au}\) and eccentricities \(e_{\mathrm{c}}< 0.8\). 
In this region, stable configurations span a wide range of companion masses, from approximately \(8\) to \(150\,\mathrm{M_{Jup}}\), and are nearly coplanar, with mutual inclinations below \(\sim 30^\circ\) (prograde) or above \(\sim 150^\circ\) (retrograde).

By contrast, many posterior samples at \(a_{\mathrm{c}} > 2~\mathrm{au}\) display large values of \(D\), indicating strongly chaotic evolution over the integration time. 
Although a few isolated regular islands remain at larger semi-major axes, they appear predominantly around mutual inclinations above \(\sim150^\circ\), corresponding to retrograde orbits, and they represent a smaller fraction of the explored posterior volume. 
The dynamical analysis therefore favors compact configurations for the candidate companion ``c'', in nearly coplanar orbits, either prograde or retrograde.

Furthermore, from the posterior stability distribution, we find that, for a companion ``c'' in the brown-dwarf/low-mass-star regime, \(18\%\) of the cases are stable. In contrast, for a planetary-mass companion, only \(2\%\) of the cases are stable. Regarding the orbital regime, we find that \(25\%\) of the configurations with \(a_{\rm c} \leq 1\,\mathrm{au}\) are stable, whereas only \(3\%\) of the configurations with \(a_{\rm c} > 1\,\mathrm{au}\) are stable.

When we combine these two conditions, we find that a planetary-mass companion ``c'' with \(a_{\rm c} \leq 1\,\mathrm{au}\) is stable in \(47\%\) of the cases, whereas for \(a_{\rm c} > 1\,\mathrm{au}\), only \(2\%\) of the cases are stable. For a companion ``c'' in the brown-dwarf/low-mass-star regime, \(25\%\) of the cases with \(a_{\rm c} \leq 1\,\mathrm{au}\) are stable, compared with \(6\%\) of the cases with \(a_{\rm c} > 1\,\mathrm{au}\).

We conclude that the candidate companion ``c'' is dynamically plausible, although the stable solutions appear to strongly prefer the substantial fraction of brown-dwarf or very-low-mass stellar companions at shorter orbital periods.

\section{Discussion}

\subsection{System architecture}

We have measured the orbit of 51~Eri~b to exceptional precision, and ruled out vast swaths of parameter space where additional companions to 51~Eri~A were previously invoked to explain the enhanced eccentricity of the planet. Simultaneously, we have attempted to assemble evidence for an inner companion, 51~Eri~``c" (or ``B") that could explain the apparent reflex motion of the host star. 

A large fraction of the phase space that a currently bound companion \textit{exterior} to 51~Eri~b could pump the planet's eccentricity via the eccentric von Zeipel-Lidov-Kozai (ZLK) effect \citep{vonZeipel1910, Lidov1962, Kozai1962, Naoz2016} within the system's lifetime has been ruled out by our JWST coronagraph imaging. It has been known for over a decade that the nearby ($a\sim2000\,\mathrm{au}$), co-moving stellar binary GJ~3305~AB \citep{Feigelson2006, Calissendorff2022} cannot induce ZLK oscillations on b to explain the eccentricity of the planet \citep{Montet2015}. \citet{Denis2026} postulated that lower mass companions ($0.2-2\,\Mjup$) at wider separations ($10-100\,\mathrm{au}$) would be invisible to current ground based observatories and could still plausibly induce the ZLK effect. We have ruled out bound companions down to $1\,\Mjup$ beyond distances of $25\,\mathrm{au}$, and to $0.25\,\Mjup$ beyond $100\,\mathrm{au}$ by observing the system with JWST coronagraphy, which leaves less phase space within which a distant perturber may induce an appreciable change of the orbital parameters of b within 20-30 Myrs \citep[see Figure E2 in][]{Denis2026}. \added{A smaller planet or series of planets could remain undetected exterior to b and could still dynamically influence it, depending on it's mass. It seems unlikely that a less massive body could pump 51~Eri~b's eccentricity all the way to 0.57 \citep[][]{Naoz2016}, but confirming this will require future numerical work, which we defer to after the DR4 data release.}

The evidence supporting a candidate inner companion, 51~Eri~``c," is only as strong as our understanding of the astrometric data. The sampler converged on almost always including an additional companion, because otherwise the absolute astrometry exhibits significant residuals when compared with the one planet orbit. This hypothesis test does not preclude a systematic source of uncertainty in the absolute astrometry, however. The evidence from the relative astrometry is not strong, as equivalent single planet trajectories can be drawn through the GRAVITY astrometry with only marginally poorly fit residuals. Because the underlying astrometric time series is as of yet unavailable, we proceed in this paper by interpreting the strong model preference for a two companion fit as suggestive evidence for a ``candidate" inner companion. 

The existence of an additional companion in the system presents a compelling explanation for the high eccentricity of the known planet, the range of possible deviations to the known planet's orbit, and the absolute astrometry of the host star, but will require additional observations to truly confirm. In the HD~206893 system, GRAVITY observations of the orbit of HD~206893~B indicated that this known companion was incapable of explaining the host star's astrometric acceleration \citep{Kammerer2021}; subsequent GRAVITY observations were able to directly detect the giant planet HD~206893~c based on two-planet orbit modeling of the known orbit and astrometric acceleration \citep{Hinkley2023}. The placement of the GRAVITY fiber can be chosen strategically to enable the detection of companions at very close separations \citep[$\ll100\,\mathrm{mas}$,][]{Pourre2024}. A similar strategy could be employed to detect 51~Eri~``c," if it exists. Most recently, the G23H catalog (or, the equivalent GaiaPMEX catalog) with the same information content as the absolute astrometry considered in this work has been used to infer the presence of a fifth planet interior to the HR~8799 system \citep{Lagrange2026, Nguyen2026}.

\begin{figure}
    \centering
    \includegraphics[width=\linewidth]{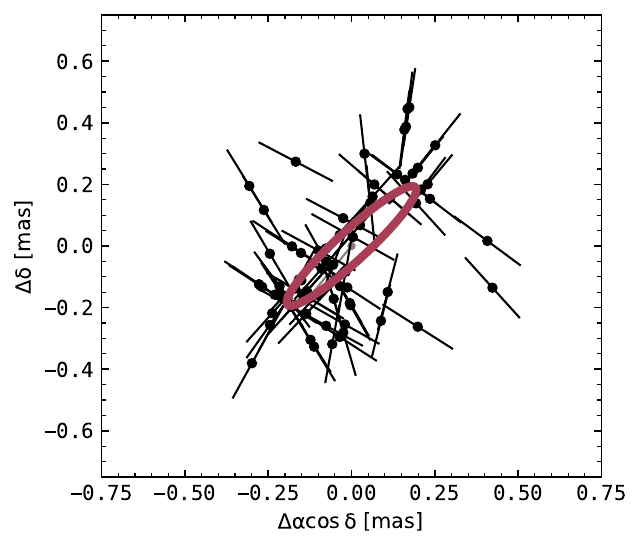}
    \caption{ Simulated epoch astrometry of 51~Eri~A after subtracting the star's proper motion and parallax assuming there is a $20\,\Mjup$ companion on a low eccentricity, 0.6~year orbit interior to 51~Eri~b. Predicted Gaia observations are shown as black scatter points with uncertainties aligned with their associated scan angles; the orbit for the candidate companion is indicated by the mauve curve, with periastron passage indicated by the gray line. The predicted averaged along-scan uncertainties for 51~Eri~A in Gaia DR4 are of order 0.1 mas, which will enable the confirmation or rejection of the candidate.}
    \label{fig:gaiamock}
\end{figure}

Luckily, the existence of such a companion should also be verified or refuted by analyzing future Gaia data releases. We used the \texttt{gaiamock}\footnote{\url{https://github.com/kareemelbadry/gaiamock}} python package \citep{El-Badry2024} to simulate the epoch astrometry that will be released in Gaia DR4. Figure \ref{fig:gaiamock} illustrates the astrometric orbit of the maximum a-posteriori sample for 51~Eri~``c" in comparison with the predicted Gaia DR4 epoch astrometry. Averaged across the multiple detectors per observation, the predicted along scan sensitivity for a G magnitude 5.14 star is of order $0.1\,\mathrm{mas}$, about a fifth the amplitude of the candidate companion's orbit. This precision will be sensitive to the vast majority of the posterior samples from our orbit fit. If no such companion exists, the epoch astrometry should indicate it clearly, enabling a confirmation or rejection of this candidate once the fourth data release is made public in December 2026. Radial velocity observations, which will require a treatment of the rapid spin and pulsations of the host star, could also place additional constraints on this putative companion. Dynamical interactions with an inner companion can also explain the low fractional infrared luminosity of the system, despite its young age \citep{Rebull2008, Riviere-Marichalar2014}, as any planetesimal belt left over after disk dispersal would likely have been disrupted as b's initially circular orbit was disrupted.

\subsection{What has excited 51~Eri~b to e=0.57?}

Previously, \citet{Balmer2025a} found the close cousin to 51~Eri~b, AF~Leporis~b ($M_{\rm b}=3.7\pm0.5\,\Mjup$, $a=9.0\pm0.2\,\mathrm{au}$, also a BPMG member) to have an effectively circular orbit using GRAVITY astrometry. Indeed, the majority of planetary mass, directly imaged planets have circular or low eccentricity orbits \citep{Bowler2020, DoO2023, Nagpal2023}; an upcoming paper will revisit this trend with the full set of observations taken during the course of the ExoGRAVITY Large Program (J. Roberts, in prep.). 51~Eri~b is the most notable outlier to this trend. 

Typically, low mass planets, unable to open significant gaps, are expected to emerge from the protoplanetary disk on roughly circular orbits because density waves launched by the Lindblad and co-rotation resonances of a non-circular orbit tend to damp the eccentricity quickly \citep[e.g.][equations 50-54 for the linear theory]{Tanaka2004}. Even if the planet is massive enough to open a gap in the protoplanetary disk, it is thought that corotation torques damp eccentricity faster than the Lindblad torques can excite it, again resulting in circular orbits \citep[e.g][]{Goldreich1980} except when the planet's mass ratio is high ($>0.003$) \citep[e.g.][]{Kley2006}. We have measured 51~Eri~b's mass ratio to be low in this work ($<0.003$). \citet{JLi2023} recently proposed a mechanism where a planet within the cavity of a transition disk with a small initial eccentricity can excite the planetary eccentricity to $0.1-0.6$ by secular resonances. To achieve $e\sim0.6$ with this mechanism, however, would require a very low mass for the planet (such that it likely would not be detected in thermal imaging), or a long disk decay timescale (which, around an early type star like 51~Eri~A, appears unlikely).

If 51~Eri~b did not form within a protoplanetary disk, on the other hand, its high eccentricity would be unsurprising, since wide orbit brown dwarf and stellar mass companions exhibit an isotropic distribution of eccentricities \citep[e.g.][]{Bowler2020}. The opacity limit for fragmenation lies near allowable masses for the planet \citep[e.g. $1-3\,\Mjup$,][]{Boyd2005, Bate2009}, but this formation pathway should encode abundances equivalent to the stellar abundances of the BPMG \citep[$\mathrm{[Fe/H]}=0.13\pm0.10$,][]{Reggiani2024}. Formation within a disk, with ready access to metal rich planetessimals, should encode elevated abundances that decrease with increasing mass \citep[e.g.][]{Thorngren2016}. The strongest evidence against a fragmentation driven formation is the enhanced atmospheric metallicity of the planet as measured from the CO$_2$ and CO absorption features seen by JWST \citep[$\mathrm{[M/H]}=0.7\pm0.2$,][]{Balmer2025a, Madurowicz2025}, which agrees well with the observed mass-metallicity trend for warm giant planets that have formed by core accretion \citep{Chachan2025}. It seems unlikely, therefore, that 51~Eri~b formed from gravitational collapse and turbulent fragmentation, a pathway which could have otherwise explained its elevated eccentricity.

It is possible that a stellar close encounter during the open cluster stage could have pulled 51~Eri~b onto a wide orbit \citep{Laughlin1998, Rodet2017}, though stellar flybys have difficulty in exciting eccentricities beyond $\Delta e\sim0.1$ except when the closest approach is of order the planet's semi-major axis \citep[e.g.][]{Zakamska2004, Li2020}. 

\subsubsection{Planet-planet scattering?}

An evocative remaining explanation for the planet's properties are dynamical interactions with other planets \citep[for a review, see e.g.][]{Davies2014}, which has motivated our hunt for a possible inner companion. Repeated crossing of resonances between planetary orbits during migration within a protoplaentary disk are thought to excite the eccentricities of the less massive planet \citep{Chiang2002}. After disk dispersal, planet-planet scattering has been posited to explain the excess of eccentric orbits of the exoplanet population \citep[e.g.][]{Ford2008}, the eccentricity excess of close-in planets orbiting metal rich stars \citep{Dawson2013}, the eccentric orbits of wide separation planets in multi-planet systems \citep[e.g.][]{BardalezGagliuffi2021, BardalezGagliuffi2025}, and the architecture of mutually inclined multi-planet systems \citep[e.g.][]{Lu2025, Zhang2025, Lu2026}. The vast majority of stable orbits in this regime are retrograde in mutual inclination (150-180$^\circ$), which would require potentially triple-planet scattering, a stellar flyby, or some other perturbation to achieve. 

Inter-related to planet-planet scattering is the process of giant impacts, or major-mergers, where before or during gas accretion planetary cores collide after chaotic interactions. \citet{Frelikh2019} show that a phase of giant impacts are consistent with the eccentricity distribution of warm Jupiters, and predict that scattering and mergers produces a population of high mass planets between 1-8 au. \citet{Ginzburg2020} argue that this process can explain the apparent atmospheric metal enrichment of super-Jupiters \citep{Wang2025, Ruffio2026, Xuan2026}, like 51~Eri~b's. Most recently, \citet{Blunt2026} presented evidence in a uniform sample of radial velocity detected planets that super-Jupiters exhibit elevated eccentricities compared to lower mass gas giants at similar separations, and \citet{Alqasim2025} showed a similar trend for warm transiting giant planets but as a function of radius rather than mass.

\par In this context, the constraints we have presented on the reflex motion of 51~Eri~A that hint at an inner companion represent a natural explanation for the high eccentricity of 51~Eri~b. If the inner companion has a planetary mass with mass $M_c\leq13\,\Mjup$, as is favored in the orbit fits assuming log-normal priors on the component masses, the system would represent an outcome of an epoch of planet-planet interactions, potentially chaotic planet-planet scattering. 

\subsubsection{Circumbinary evolution?}

Outside of critical separations close to the path of the binary orbit, planet formation in circumbinary disks proceeds relatively robustly, very similar to formation around single stars \citep[e.g.][]{Bromley2015}. If the putative inner companion 51~Eri~``c" has a brown-dwarf or stellar mass with $M_c\geq13\,\Mjup$, as is favored in the orbit fits assuming uniform priors on the component masses, the dynamical history of 51~Eri~b could be better understood as an outcome of circumbinary planet formation. In our stability analysis, the vast majority of eccentricities, masses, and mutual inclinations in this circumbinary regime of the orbital posterior distribution appear stable in the long-term, allowing for stable pro- and retrograde P-type orbits, except for mutual inclinations along an instability corridor from roughly 90-120$^\circ$. 

In hydrodynamic simulations of circumbinary giant planet formation, the evolution of Jupiter mass protoplanets is typically dictated by the protoplanet's interaction with the 4:1 resonance of the binary orbit \citep{Nelson2003,Pierens2008}. The resonant interaction drives up the eccentricity of the planet until it scatters off the smaller binary component, either being ejected or surviving on a wider separation orbit. Then, torques from the circumbinary disk drive the outwards migration of the planet. The protoplanet can also stall just outside the 4:1 resonance, entering a stable orbit as the disk disperses, or the eccentricity of the binary can drive an eccentricity in the disk, which can also stall the inward migration of the protoplanet. In samples of the orbit posteriors with masses reflective of a central binary, the binary has separations $0.1{-}1\,\mathrm{au}$, which have 4:1 resonances from $0.4{-}4\,\mathrm{au}$. As discussed above, 51~Eri~b has a periastron passage of $3.98\pm0.03\,\mathrm{au}$ given its eccentric orbit, and therefore the samples are inclusive of orbits that could have reasonably been in 4:1 resonance with 51~Eri~b in the past. The interaction with the 4:1 resonance appears to be more frequently stable for low mass ratio binaries $q\lesssim0.1$ \citep{Nelson2003}, which is also consistent with our sampled orbits ($q=0.015\pm0.015$ when $M_c\geq13\,\Mjup$).

\section{Conclusion}

We have measured the orbit of the directly imaged planet 51~Eri~b to exceptionally high precision using dual-field optical interferometry at the VLTI. These observations unambiguously confirm that the planet has a high eccentricity orbit with $e_b=0.573\pm0.005$, $a_b=9.35\pm0.04\,\mathrm{au}$, and $i_b=152.2\pm0.5^\circ$. We have also analyzed deep space-based 4.4\um coronagraphic observations of the system from JWST/NIRCam, which rules out planets more massive than Jupiter beyond $25\,\mathrm{au}$ and more massive than Saturn beyond $100\,\mathrm{au}$.

We explored the potential for a third-planet in the system to explain the absolute astrometry of the star and any residual motion observed on the outer planet's orbit. We found plausible solutions with $M_c\in[1.0, 2900]_{95\%}\,M_{\rm Jup}$ and $a_c=[0.3, 3.1]_{95\%}\,\mathrm{au}$. Samples with $a_c\lesssim1\,\mathrm{au}$ appear most likely to be long-term stable according to our stability analysis. If this candidate is astrophysical, it should be apparent in Gaia DR4 epoch astrometry even under conservative assumptions of bright star astrometric performance. Additional GRAVITY+ observations of the known planet targeting a 3-6 month cadence will also be sensitive to the reflex motion of 51~Eri~A due to this putative companion and will help refine the mass and separation of this companion. 51~Eri~c may be amenable to direct detection with GRAVITY+, in dual-field on-axis observations \citep{Pourre2024}, or in classical interferometric observables like the visibilities and closure phases. To achieve sensitivity to sub-stellar companions would likely require on-star observations with a dedicated interferometric calibrator, which the on-star observations we have presented here do not have, but which could be obtained in a more classic interferometric observing SCI-CAL sequence. Stellar obliquity measurements of the host star with optical interferometry (say, from CHARA or the VLTI) will be crucial to interpreting both the formation and architecture of the 51~Eri planetary system, and the long term stability of the system. 

Planet-planet scattering and circumbinary planet formation undoubtedly shape the observed distribution of close-in exoplanets, but their impacts on the architectures of widely separated planets is still uncertain. The 51~Eri system presents an exciting opportunity to further study the chaotic interactions that can dramatically reshape the evolution of giant planets.

\begin{acknowledgements}
Thanks to Anne-Lise Maire for her contributions to this project during the ExoGRAVITY Large Program. Thanks as well to Linling Shuai, Bin Ren, and Tiger Lu for helpful discussions regarding stellar flybys. We thank the anonymous reviewer for their detailed report, which improved the presentation and analysis of the paper substantially. 

W.O.B. is grateful for support from the Heising-Simons Foundation 51 Pegasi b Fellowship (grant \#2026-6639).
Some of this work was carried out by W.O.B. at the Advanced Research Computing at Hopkins (ARCH) core facility (rockfish.jhu.edu), which is supported by the National Science Foundation (NSF) grant number OAC1920103.

J.W.X is grateful for support from the Heising-Simons Foundation 51 Pegasi b Fellowship (grant \#2025-5887).

A.C.M.C, P.J.V.G., R.P.S. acknowledge support from FCT - Funda\c{c}\~ao para a Ci\^encia e a Tecnologia, I.P., Portugal, in the framework of the projects
2024.01252.BD, CFisUC UID/04564/2025 (\href{https://doi.org/10.54499/UID/04564/2025}{10.54499/UID/04564/2025}), and Center for Astrophysics and Gravitation (CENTRA/IST/ULisboa) UID/PRR/00099/2025 (\href{https://doi.org/10.54499/UID/PRR/00099/2025}{10.54499/UID/PRR/00099/2025}) and UID/00099/2025 (\href{https://doi.org/10.54499/UID/00099/2025}{10.54499/UID/00099/2025}).
The stability maps were performed at the Oblivion Supercomputer at the University of \'Evora (\href{https://oblivion.hpc.uevora.pt}{oblivion.hpc.uevora.pt}).
\end{acknowledgements}

\begin{contribution}
    W.O.B. led the project and wrote the manuscript. L.P. advised W.O.B. and guided the development of the project. W.O.B., S.L., A.-L.M., and M.D.P. led the proposals to obtain the data. W.O.B., S.L., and M.N. processed the data. W.O.B. and W.T. fit the orbits. W.T. and D.B. developed key aspects of the orbit fitting code. R.P.S., A.C.M.C., and P.J.V.G. undertook the dynamical stability analysis. All other authors contributed to the ExoGRAVITY or JWST TST-GTO  collaboration, including attending observing runs, developing preliminary results, providing feedback, and contributing edits to the manuscript.
\end{contribution}

\facilities{VLTI(GRAVITY), VLT(SPHERE, HiRISE), Gemini(GPI), Keck(NIRC2), JWST(NIRCam), Hipparcos, Gaia}

\software{Python \citep{python}, 
Julia \citep{Julia2017}, 
\texttt{AstroPy} \citep{astropy:2013, astropy:2018, astropy:2022}, 
\texttt{NumPy} \citep{harris2020array},  
\texttt{Octofitter.jl} \citep{Thompson2023}, \texttt{PairPlots.jl} \citep{pairplots}, 
\texttt{ParetoSmooth.jl} \citep{ParetoSmooth.jl}
\texttt{Pigeons.jl} \citep{Surjanovic2023}, 
\texttt{pyKLIP} \citep{Wang2015}, 
\texttt{SciPy} \citep{2020SciPy-NMeth},
}

\bibliography{ceri}{}
\bibliographystyle{aasjournal} 

\global\suppressAffiliationsfalse
\allauthors
\end{document}